# The local similarity theory, including the "spectral" Prandtl mixing length. Dissipation of energies and structural parameters in the atmospheric CBL.

A. N. Vulfson [a] [b], P. V. Nikolaev [c]

[a] Water Problems Institute, Russian Academy of Sciences, Gubkina str., 3, Moscow, Russia, 117971,
email: vulfson@iwp.ru

[b] National Research University Higher School of Economics, Myasnitskaya st., 20, Moscow, Russia, 101000

[c] School of Engineering, The University of Manchester, Oxford rd., Manchester, UK, M13 9PL,
email: petr.nikolaev@manchester.ac.uk

**Abstract**

The paper discusses a variant of the local similarity theory, employing the second moment of vertical velocity and the "spectral" Prandtl mixing length as basic parameters. This approach allows expressing the turbulent exchange coefficient, dissipations of kinetic energy and buoyancy square, mixed moments of the buoyancy and vertical velocity, as well as structural parameters solely through two independent basic parameters of local similarity. Comparison of local similarity approximations with experimental data convincingly confirms the correctness of the proposed relationships and demonstrates the high efficiency of the considered theory. It is established that the approximations of the turbulent exchange coefficient and dissipations of local similarity are fully consistent with the semi-empirical theories of Prandtl, Hanna, Richardson-Onsager and the Kolmogorov-Obukhov spectral theory with the law -5/3 for the atmospheric CBL. Thus, the new theory of local similarity significantly extends the scope of applications of well-known turbulence theories and substantially complements our understanding of turbulent convection.

**Key words**

Convective boundary layer. Theory of local similarity. "Spectral" mixing length. Semi-empirical Prandtl's theory of turbulence.

# 1. Introduction

The theory of similarity and dimensional analysis finds broad applications in turbulence theory. At the basis of dimensional theory lies the well-known Buckingham theorem (Buckingham 1914; Barenblatt 1996). Modern mathematical justification of the similarity theory relies on the theory of Lie groups (Olver 1993). Therefore, under conditions corresponding to experimental data, the results of similarity theory are self-sufficient and do not require additional justification.

Hereafter, we will consider the mean turbulent parameters in the developed atmospheric convective boundary layer (CBL) above a horizontal homogeneous underlying surface under weak wind conditions. Let $z$ be the height above the horizontal underlying surface and $h$ be the height of the convective layer. In the developed convective layer, averaging meteorological parameters over the area allows constructing vertical profiles of turbulent moments depending on the dimensionless height $z/h$.

Classical similarity theory is undoubtedly one of the most effective methods for constructing analytical approximations of turbulent moments. In 1953, A.S. Monin and A.M. Obukhov first used this theory to approximate turbulent moments in the surface layer of the atmosphere $0 \leq z/h \leq 0.1$ (Monin and Obukhov 1953, 1954).

The description of the turbulent moments of the windless convective layer $0 \leq z/h \leq 1$ can also be implemented within the framework of local similarity theory. This theory presupposes the a priori specification of several basic dimensional parameters characterizing turbulent mixing at each dimensionless height $z/h$. In this case, approximations of turbulent moments are built based on dimensional theory in the form of generalized monomials depending on the selected basic parameters.

For the atmospheric CBL without wind shear, the theory of local similarity was first proposed in the work by Zeman and Lumley (1976). This approach uses height $z$ and local buoyancy flux $gs_\theta$ as basic parameters, analogous to the Monin-Obukhov similarity theory (MOST) for the windless surface layer. Various modifications of the theory by Zeman and Lumley (1976) were discussed in studies (Sorbjan 1986, 1987, 1988, 1990, 1991; Vulfson et al. 2004). All these modifications use the same basic parameters. It can be shown that modified versions of local similarity correspond to experimental data in the lower half of the atmospheric convective layer $0 \leq z/h \leq 0.5$, for more details see (Vulfson and Nikolaev 2022).

A specific variant of local similarity theory, using the semi-empirical Prandtl's theory, was proposed by Vulfson and Nikolaev (2022). It is known that, according to Prandtl's theory, for the coefficient of turbulent heat exchange $K_H$ and the dissipation of kinetic energy $\varepsilon_b$, the following relationships hold: $K_H = l_P \cdot \sqrt{\overline{w^2}}$ and $\varepsilon_b = \lambda_{\varepsilon b} l_P^{-1} \left(\overline{w^2}\right)^{3/2}$, where $\overline{w^2}$ is the second moment of vertical velocity; $l_P$ is the mixing length, and $\lambda_{\varepsilon b} > 0$ is a positive constant. Therefore, the Prandtl's theory relationships allow us to assume that instead of the similarity parameters $z$ and $gs_\theta$, we can use the basic parameters $\overline{w^2}$ and $l_P$.

In classical Prandtl's theory, the mixing length $l_P$ is a quantity that is not clearly defined. That is why there is a wide set of a priori relations to specify $l_P$, see, for example, (Holt and Raman 1988; Peña et al. 2010). The variant of the theory by Vulfson and Nikolaev (2022) assumes that $l_P = l_{PS}$, where $l_{PS}$ is the "spectral" mixing length. Moreover, it is assumed that $l_{PS}$ is proportional to the integral scale of turbulence $\Lambda_{mw}$, as specified by Hanna (1968) and Caughey and Palmer (1979). Such an approach eliminates the uncertainty of definition $l_P$ and gives physical meaning to the concept of mixing length.

The new theory of local similarity (NTLS), a variant of which considered by Vulfson and Nikolaev (2022), allows describe a broad class of mixed turbulent moments of 'adiabatic' buoyancy and vertical velocity $\overline{(g\theta_a)^m w^n}$, where $m \geq 0$ and $n \geq 0$ are non-negative integers and the bar symbol indicates horizontal surface averaging. A clear representation of this class of turbulent moments can be obtained during laboratory experiments in a water tank. The theory allows expressing a wide class of turbulent moments of the convective layer through the basic parameters $\overline{w^2}$ and $l_{PS}$. Thus, among two dozen different vertical profiles of turbulent moments, only two profiles are considered independent. This circumstance is a fundamental step in the structural description of turbulent convection.

It should be emphasized that the approximations of the NTLS are valid in the layer $0 \leq z/h \leq 0.7$ corresponding to the location of the ensemble of small convective thermals. Whereas known variants of local similarity theory correspond to observation data in a thinner layer $0 \leq z/h \leq 0.5$. This means that the new variant of the theory is more efficient for applications.

In the subsequent articles by Vulfson and Nikolaev (2024a,b), an additional class of mixed turbulent moments of Boussinesq buoyancy and vertical velocity $\overline{(g\theta_b)^m w^n}$ was considered. Clear representations of this class of turbulent moments can be obtained through field measurements in the atmosphere and ocean. The version of the local similarity theory also allows expressing this class of turbulent moments through the basic parameters $\overline{w^2}$ and $l_{PS}$.

The local similarity theory relationships always contain uncertain parameters, the values of which must be determined from additional a priori assumptions. In the proposed theory, the basic parameters $\overline{w^2}$ and $l_{PS}$ are chosen in such a way that in the surface layer $0 < z/h < 0.1$, the approximations of turbulent moments of the local similarity theory are identical to the moments of the MOST. This circumstance allows estimating the uncertain parameters of local similarity theory from known measurements in the surface layer. The identity of local similarity approximations and MOST approximations in the surface layer allows considering the proposed theory of local similarity as a generalization of the Monin-Obukhov theory under conditions of free convection.

In the present study, a significant addition to the papers (Vulfson and Nikolaev 2022, 2024a,b) is provided. The proposed study examines new approximations:

- of dissipations of kinetic energy and buoyancy squared, including the spectral mixing length,

- of relationships for the structural parameters of velocity and buoyancy of the second and third order, consistent with observation data.

Additionally, the alignment of NTLS with the results of field experiments Minnesota-1973, Ashchurch-1976, Beauce-1976, Toulouse-1976, IHOP_2002, AVEC-2006 and ShUREX-2016, has been completed. Furthermore, comparisons are made between NTLS relationships and approximations of semi-empirical theories by Prandtl, Hanna, and Richardson-Onsager as well as the Kolmogorov-Obukhov spectral theory with the law -5/3. The results of comparing not only confirm the validity of the proposed approximations but also convincingly demonstrate the effectiveness and advantages of NTLS compared to already known theories.

The paper is organized as follows. In Section 2, a modern scheme of the vertical structure of the atmospheric convective layer is provided. The concepts of "adiabatic" and "Boussinesq" buoyancy are defined. Modified forms of convection theory equations in a static atmosphere, employing buoyancy as a new variable, are presented. In Section 3, the "spectral" mixing length and the second moment of vertical velocity are defined as the basic parameters of NTLS. An approximation of the "spectral" mixing length, chosen to be proportional to the integral scale of turbulence, is considered. A special approximation is chosen for the second moment of vertical velocity. Two-parameter approximations of the mixed moments of buoyancy and vertical velocity are constructed. In Section 4, comparisons with the semi-empirical Prandtl's turbulence theory are made. Approximations of the turbulent eddy heat exchange coefficient and turbulent energy dissipation are considered. In Section 5, comparisons with Hanna's semi-empirical turbulence theory are made. In Section 6, comparisons with Richardson's semi-empirical turbulence theory are made. In Section 7, approximations of the structural parameters of velocity and buoyancy of the second and third order are presented. Approximations of structural parameters of classical and local similarity theories are compared. In Section 8, comparisons with the spectral Kolmogorov-Obukhov turbulence theory with the laws -5/3 for the convective layer are made. Section 9 presents the main conclusions of the study.

## 2. Mathematical description of turbulent convection in the atmospheric boundary layer

Let us consider systems of equations and boundary conditions that implement the hydrodynamic description of turbulent convection in the atmospheric boundary layer.

### 2.1. Vertical structure of the convective layer

Suppose that $x$, $y$, $z$ are the coordinates of the Cartesian coordinate system, in which the $z$-axis is directed opposite to the acceleration of gravity $g$, and the $x$ and $y$-axes are located on a flat homogeneous surface; $\mathbf{u}(x, y, z, t)$ and $w(x, y, z, t)$ are the components of the velocity vector along the $xy$ plane and the $z$-axis, respectively.

Let $\Theta(x, y, z, t)$ be the local value of potential temperature, $\overline{\Theta}(z)$ be the background value of potential temperature, $\Theta_0 = const$ be the mean value of potential temperature $\overline{\Theta}(z)$ over the entire convective layer, and $d\overline{\Theta}/dz$ be the temperature stratification of the atmosphere. The value $\Theta'_b(x, y, z, t) = \Theta(x, y, z, t) - \overline{\Theta}(z)$ represents the local fluctuation of potential temperature in the stratified atmosphere.

Following Turner (1962), we introduce the local "Boussinesq" buoyancy:

$$g\theta_b(x, y, z, t) = g\Theta'_b(x, y, z, t)/\Theta_0 = g\{\Theta(x, y, z, t) - \overline{\Theta}(z)\}/\Theta_0 \qquad (1)$$

The magnitude of $g\theta_b$ can be determined based on field measurements of temperature in the CBL.

Obviously, $g\theta_b w$ represents the local vertical flux of "Boussinesq" buoyancy. Let $gs_\theta(z) = \overline{g\theta_b w}$ be the average vertical flux of "Boussinesq" buoyancy at an arbitrary level $z$, the dimension of which is $[gs_\theta] = m^2/s^3$. Hereafter, the overline denotes horizontal averaging. It is also assumed that $gs_\theta(0) = gS_\theta$, where $gS_\theta$ is the heat flux from the underlaying surface

Next, we will use a well-known CBL model used, for example, in the studies (Sorbjan 1988; Fedorovich and Mironov 1995; Gentine et al. 2013), and presented in Fig. 1.

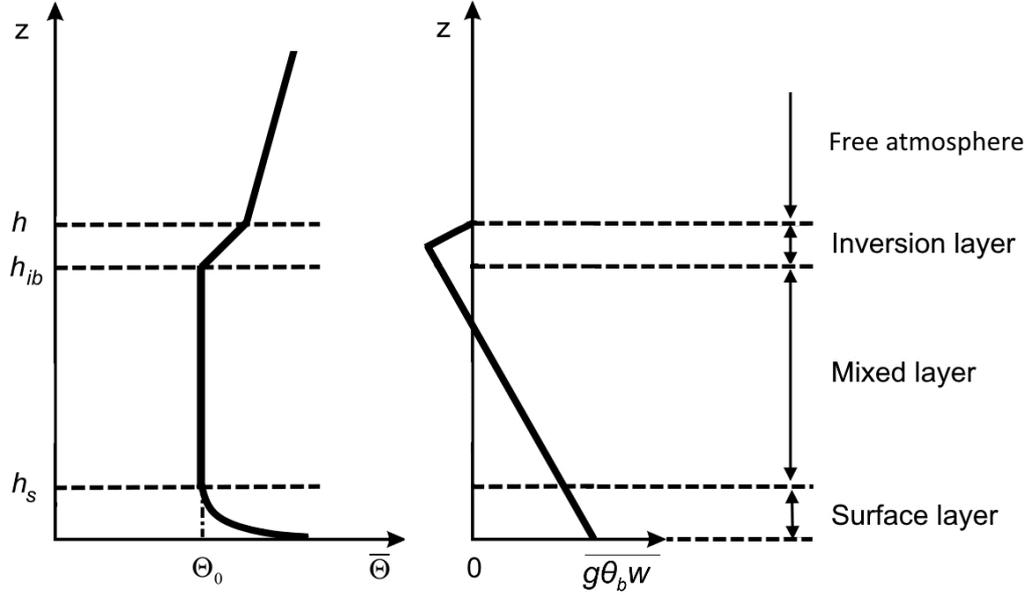

**Fig 1.** The profiles of the potential temperature and the mean buoyancy flux in the convective boundary layer of the atmosphere (Fedorovich and Mironov 1995)

We define the height of the convective layer $h$ as the minimum level of the free atmosphere, at which the gradient of potential temperature reaches its maximum local positive value, $d\overline{\Theta}/dz\big|_h = \max_z(d\overline{\Theta}/dz)$. For more details on methods for determining the height of the convective layer, see (Sullivan et al. 1998).

In the proposed model, the convective boundary layer of the atmosphere with a height $h$ consists of three horizontal tiers.

The tier $h_{ib} < z < h$ is called the inversion layer. Its lower boundary is located at the level $h_{ib} \approx 0.85h$. The atmospheric stratification in the inversion layer is stable. It is noted that according to Boers (1989), the thickness of the inversion layer $h - h_{ib}$ depends on the Richardson number $Ri$.

The tier $h_s < z < h_{ib}$ is called the mixed layer. Its lower boundary is located at the level $h_s \approx 0.1h$. The atmospheric stratification in the mixed layer is neutral, $d\overline{\Theta}/dz = 0$, and the average potential temperature is nearly constant, $\overline{\Theta}(z) = \Theta_0 = const$. Since this layer has the greatest vertical extent, it can be considered that, in the entire CBL, it is $\frac{1}{h}\int_0^h \overline{\Theta}(z)dz \approx \Theta_0$. The average buoyancy flux $gs_\theta(z) = \overline{g\theta_b w} > 0$ is positive in the most part of the mixed layer.

The tier $0 < z < h_s$, that is adjacent to the underlying surface $z = 0$, is called the surface layer. In the surface convective layer, the atmospheric stratification is unstable $d\overline{\Theta}/dz \leq 0$, but the heat flux is nearly constant, $gs_\theta(z) \square gS_\theta$, according to (Monin and Obukhov 1954). Equality $\overline{\Theta}(h_s) = \Theta_0$ should be considered as the equation for determining the height of the convective surface layer $h_s$, see Figure 1, for more details see (Kaimal et al. 1976).

Suppose $\Theta'_a(x,y,z,t) = \Theta(x,y,z,t) - \Theta_0$ is the local fluctuation of potential temperature in a neutrally stratified atmosphere. By analogy with (1), we can introduce the local "adiabatic" buoyancy $g\theta_a(x,y,z,t)$, assuming that:

$$g\theta_a(x,y,z,t) = g\Theta'_a(x,y,z,t)/\Theta_0 = g\{\Theta(x,y,z,t) - \Theta_0\}/\Theta_0 \qquad (2)$$

Comparison of (1) and (2) shows that the "adiabatic" and "Boussinesq" buoyancy are practically identical in the mixed layer but significantly differ in the surface and inversion layers.

Note that $g\theta_a$ can be observed by measuring the temperature in a water tank with homogeneous liquid. Buoyancy measured in laboratory experiments with homogeneous liquid is also called "Archimedean" buoyancy.

The convective layer structure, schematically represented in Figure 1, uses the classical approximation (Wyngaard et al. 1978), according to which:

$$\frac{g}{\Theta_0}\frac{\partial}{\partial z}\overline{\Theta} = \frac{g}{\Theta_0}\frac{\partial}{\partial z}(\overline{\Theta} - \Theta_0) = \frac{\partial}{\partial z}\overline{g\theta_a} = \begin{cases} 0, & \text{if } h_s/h \leq z/h \leq h_{ib}/h \\ -\lambda_\theta^o (gS_\theta)^{2/3} z^{-4/3}, & \text{if } 0 < z/h < h_s/h \end{cases} \qquad (3)$$

The first relation (3) indicates neutral temperature stratification of the mixed layer, $h_s \leq z \leq h_{ib}$. The second relation (3) corresponds to the free-convective limit of the Monin-Obukhov similarity theory in the surface layer. The coefficient $\lambda_\theta^o = 0.93$ in relation (3) can be chosen according to measurements by Kader and Yaglom (1990) and Kramm et al. (2013).

It can be shown that at an arbitrary level $z$, the average vertical fluxes of "Boussinesq" and "adiabatic" buoyancy coincide (Vulfson and Nikolaev 2024a):

$$gs_\theta(z) = \overline{g\theta_b w} = \overline{g\theta_a w} \qquad (4)$$

Equality (4) is called the buoyancy flux invariance condition.

The existence of parameters $gS_\theta$ and $h$ allows introducing modified Deardorff parameters for velocity and buoyancy in the convective layer (Deardorff 1970):

$$w_D = h^{1/3}(gS_\theta)^{1/3}, \quad g\theta_D = h^{-1/3}(gS_\theta)^{2/3} \qquad (5)$$

The Deardorff parameters (5) are convenient for normalizing average turbulent moments.

## 2.2. Modified forms of convection equations under no wind conditions

The hydrodynamic description of turbulent processes in a fluid relies on two fundamentally different approaches. The first approach is implemented within the framework of the Boussinesq convection equations for stratified fluid. The mathematical foundation of this system was proposed by Spiegel and Veronis (1960). The Boussinesq equations are specifically aimed at describing oceanic turbulence. The second approach to describing turbulent processes is implemented within the framework of equations proposed by Mihaljan (1962) for homogeneous fluid. These equations are directed towards laboratory descriptions of turbulent convection in water tanks. A detailed discussion of these approaches is provided in the monograph by Vallis (2017).

Analogues of these approaches also exist for describing turbulent processes in the atmosphere. The first approach is realized within the framework of the Boussinesq convection equations for stratified atmosphere. In this equation system, the background potential temperature of the air $\bar{\Theta}$ is non-uniform and varies with height, $\bar{\Theta} = \bar{\Theta}(z)$. The mathematical foundation of the Boussinesq system for atmospheric layers of thickness $h \simeq 1-2\ km$ follows from the studies by Vulfson (1981) and Mahrt (1986).

The second approach is implemented within the framework of equations proposed by Ogura and Phillips (1962). In this system of equations, the background potential temperature of the air is uniform and does not vary with height, $\Theta_0 = const$. Similar equations were first proposed by Batchelor (1953). It should be emphasized that for turbulent atmosphere, equations practically identical to the system proposed by Ogura and Phillips (1962) were also used in the original study by Monin and Obukhov (1954).

Next, we will consider the convective layer of a stratified atmosphere, $\bar{\Theta} = \bar{\Theta}(z)$. Let $\mathbf{u}(x,y,z,t)$ be the component of the velocity vector along the plane $xy$; $\mathbf{U}(z)$ be the mean wind speed; $\mathbf{u'} = \mathbf{u} - \mathbf{U}(z)$ be the deviation of the local horizontal velocity from the mean wind $\mathbf{U}(z)$. Under conditions of free convection in the absence of wind or forced convection with weak wind, it can be assumed that $\mathbf{U}(z) = \mathbf{0}$. Let $w(x,y,z,t)$ be the component of the velocity vector along the $z$ axis and $\bar{w} = 0$ the mean vertical velocity. The values $\rho(x,y,z,t)$ and $p(x,y,z,t)$ represent the local values of density and pressure in the atmosphere; respectively; $\bar{\rho}(z)$ and $\bar{p}(z)$ are the background values of density and pressure in the stratified atmosphere, related by the equations of state of ideal gas and statics.

Let $\bar{\rho}_a(z)$ be the background value of density of a static adiabatic atmosphere; $\bar{\rho}_0 = \bar{\rho}_a(0)$ be constant value of average density at the underlying surface; and $p'(x,y,z,t) = p(x,y,z,t) - \bar{p}(z)$ be the deviation of the local pressure $p(x,y,z,t)$ from its static background value $\bar{p}(z)$; $\Phi_b = p'/\bar{\rho}_a \approx p'/\bar{\rho}_0$ be the modified pressure disturbance in a stratified atmosphere.

Next, we will use the modified Boussinesq equation system, including "Boussinesq" buoyancy (1), then

$$\begin{cases} \dfrac{d}{dt}\mathbf{u'} = -\nabla_h \Phi_b, & \dfrac{d}{dt}w = -\dfrac{\partial}{\partial z}\Phi_b + (g\theta_b) \\ \dfrac{d}{dt}(g\theta_b) + g\Gamma w = 0, & \nabla_h \cdot \mathbf{u'} + \dfrac{\partial}{\partial z}w = 0 \end{cases} \quad (6)$$

where $g\Gamma = g\Theta_0^{-1}\partial\bar{\Theta}(z)/\partial z = \partial \overline{g\theta_a}/\partial z$ is the stratification parameter; $\nabla_h = \mathbf{i}\partial/\partial x + \mathbf{j}\partial/\partial y$ is the horizontal Hamiltonian operator; $\mathbf{i}$ and $\mathbf{j}$ are the unit vectors along the $x$ and $y$ axis respectively.

Taking into account the adopted notation, the stratification parameter (3) will take the form:

$$g\Gamma = \dfrac{\partial}{\partial z}\overline{g\theta_a} = \begin{cases} 0 & \text{if } h_s \leq z \leq h_{ib} \\ -\lambda_\theta^o (gS_\theta)^{2/3} z^{-4/3}, & \text{if } 0 < z < h_s \end{cases} \quad (7)$$

The Boussinesq equations (6) is considered in a convective layer of the atmosphere with finite thickness $\Omega = \{-l_{xy} < x, y < l_{xy},\ 0 < z < h\}$, where $h \sim 1-2\ km$ and $l_{xy} \geq 5\ km$ are the height and horizontal extent of the convective layer.

Averaging the boundary conditions of system (6) at the horizontal boundaries leads to the following equalities:

$$\begin{cases} \left\{\overline{(w\mathbf{u'}\cdot w\mathbf{u'})\big|_h}\right\}^{1/2} = 0, & \overline{g\theta_b w}\big|_h = 0 \\ \left\{\overline{(w\mathbf{u'}\cdot w\mathbf{u'})\big|_0}\right\}^{1/2} = U_*, & \overline{g\theta_b w}\big|_0 = gS_\theta \end{cases} \quad (8)$$

where $U_*$ is the frictional velocity. It can be assumed that under conditions of free convection and forced convection with weak wind, $U_* = 0$.

In the case of neutrally stratified (adiabatic) atmosphere, $\bar{\Theta}(z) = \Theta_0 = const$, so in system (6) we should set $g\Gamma = 0$. Let us replace variables $g\theta_b$ and $\Phi_b$ in equation (6) with variables $g\theta_a$ and $\Phi_a$, where the modified pressure in the neutral atmosphere $\Phi_a$, defined similarly to $\Phi_b$. Then

$$\begin{cases} \dfrac{d}{dt}\mathbf{u'} = -\nabla_h \Phi_a, & \dfrac{d}{dt}w = -\dfrac{\partial}{\partial z}\Phi_a + (g\theta_a) \\ \dfrac{d}{dt}(g\theta_a) = 0, & \nabla_h \cdot \mathbf{u'} + \dfrac{\partial}{\partial z}w = 0 \end{cases} \quad (9)$$

The system of equations (9) represents the modified system of equations by Ogura and Phillips (1962), including "adiabatic" buoyancy (2). System (9) employs the continuity equation in an incompressible form and is therefore valid in a relatively thin layer $h \sim 1-2\ km$.

The condition of buoyancy flux invariance (4) allows us to transform the boundary conditions of system (6) into the boundary conditions of system (9), such that

$$\left\{ \overline{(w\mathbf{u'} \cdot w\mathbf{u'})}\big|_h \right\}^{1/2} = 0, \quad \overline{g\theta_a w}\big|_h = 0$$
$$\left\{ \overline{(w\mathbf{u'} \cdot w\mathbf{u'})}\big|_0 \right\}^{1/2} = U_*, \quad \overline{g\theta_a w}\big|_0 = gS_\theta \tag{10}$$

It should be noted that when using stratification (7), the systems of the atmospheric convection (6) and (9) are equivalent in the mixed layer $h_s \leq z \leq h_{ib}$, but differ in the surface layer $0 < z < h_s$ and the inversion layer $h_{ib} \leq z \leq h$.

The comparison of the variables of (6) and (9) with taking (4) into account gives:

$$\begin{cases} g\theta_b = g\theta_a - \int_{h_s}^{z} g\Gamma(\xi)d\xi, \quad -\dfrac{\partial}{\partial z}\Phi_b = -\dfrac{\partial}{\partial z}\Phi_a + \int_{h_s}^{z} g\Gamma(\xi)d\xi \\ \overline{g\theta_b w} = \overline{g\theta_a w} \\ \overline{g\theta_b w} = -K_H \dfrac{\partial}{\partial z}\overline{g\theta_a} \end{cases} \tag{11}$$

where $K_H$ is the coefficient of turbulent heat exchange.

The relations in the first line of equation (11) connect "Boussinesq" buoyancy quantities and modified pressure to their "adiabatic" counterparts. The correctness of the equations in the first line can be established by directly substituting them into the corresponding equations, see (Vulfson 1981). The relationship in the second line of equation (11) represents the condition of buoyancy flux invariance, see (Vulfson and Nikolaev 2024a). The relationship in the third line of equation (11) represents the well-known Boussinesq hypothesis, see (Boussinesq 1870).

It is essential to emphasize that the modified forms of the convection equation systems (6) and (9) do not contain the buoyancy parameter $g/\Theta_0$, and thus are fundamentally different from the classical forms by Spiegel and Veronis (1960) and Ogura and Phillips (1962). The fewer basic parameters in systems (6) and (9) provide a rigorous justification for applying the theory of similarity and Buckingham's theorem for approximations of turbulent moments, see (Buckingham 1914; Barenblatt 1996). The question of using classical similarity theory to approximate turbulent moments was extensively discussed in (Vulfson and Nikolaev 2024a).

It is worth noting that the modified form of the convection equations (6) - (8) was first proposed by Vulfson et al. (2004), and has also been used in the following works: (Garcia 2014; Fedorovich et al. 2017; Fodor and Mellado 2020; Vulfson and Nikolaev 2022, 2024a).

## 3. New Theory of Local Similarity: Basic Parameters and Turbulent Moments

In classical Prandtl theory, the coefficient of turbulent heat exchange and the dissipation of turbulent kinetic energy are expressed in terms of the mixing length $l_P$ and the second moment of vertical velocity $\overline{w^2}$. Here, we consider a variant of local similarity theory proposed by Vulfson and Nikolaev (2022, 2024a,b), where the "spectral" mixing length of Prandtl is chosen as a basic parameter along with the second moment of vertical velocity. It is shown that such a choice of basic parameters allows for the construction of effective approximations for moments of buoyancy and vertical velocity of any order. These relationships are significant additions to Prandtl's theory.

### 3.1. "Spectral" Mixing Length of Prandtl as a Basic Parameter

Classical Prandtl theory defines the mixing length only in the vicinity of a solid horizontal surface following the dimensional theory as $l_P \sim z$, see (Prandtl 1925). Far from the wall, the mixing length $l_P$ is not precisely defined. Therefore, various a priori relationships exist for defining the mixing length, as seen in (Holt and Raman 1988; Peña et al. 2010). Various approximations for the mixing length in modern numerical models are provided in works such as (Abdella and Mcfarlane 1997).

Several theoretical approaches exist for constructing the mixing length, including the "profile" method by Karman, as described in (von Kármán 1930; Menter and Egorov 2010), as well as the method by Lenschow and Stankov (1986), which employs the self-correlation function.

In the discussed variant of local similarity theory, the mixing length of Prandtl $l_P$ is defined based on the spectral theory of turbulence. Let $k$ be the wavenumber of the spectral decomposition of the square of vertical velocity $w^2$ at some level $z/h$; $P_w(k)$ be the spectral density of the decomposition of vertical velocity squared $w^2$, and $kP_w(k)/w_D^2$ be the normalized spectral density of vertical velocity at some level $z/h$, dependent on $kh$.

The dependence of $kP_w(k)/w_D^2$ on the dimensionless wavenumber $kh$ at different heights $z/h$, according to measurements by Lothon et al. (2009), is depicted in Figure 2. The inertial part of the spectrum is located in the region of short waves $k \gg 1$.

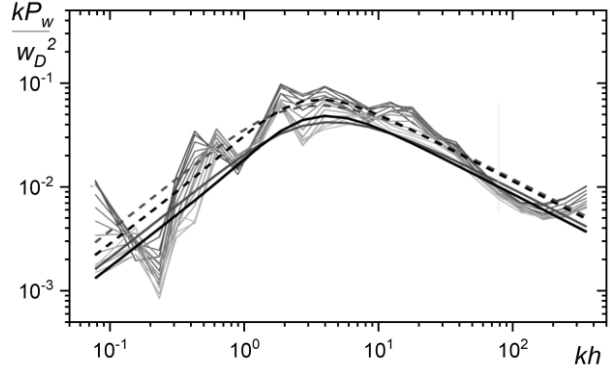

**Fig. 2** The dependence of the normalized spectral density $kP_w(k)/w_D^2$ on the wave number $kh$ at a height $z/h$. The measurements were performed on August 20, 1996, and are presented by grey lines which correspond to the different values of $z/h$ within the CBL. The solids and dashed lines are the approximations by Lothon et al. (2009).

Let $\lambda = 2\pi/k$ be the wavelength corresponding to the wavenumber $k$. According to measurements at a fixed level $z/h$, the experimental profile $kP_w/w_D^2$ reaches its maximum at a certain wavelength $\lambda = \Lambda_{mw}$ as shown in Fig. 2. This wavelength $\Lambda_{mw} = \Lambda_{mw}(z/h)$, which realizes the maximum in the spectrum of the vertical velocity field $kP_w(k)/w_D^2$, is referred to as the extreme wavelength. The magnitude $\Lambda_{mw}$ corresponds to the size of turbulent eddies with the highest kinetic energy of vertical motion. The spectral definition, $\Lambda_{mw}$, was first proposed by Hanna (1968).

Experimental and computed values of $\Lambda_{mw}/h$ are typically approximated by exponential relationships, see (Kaimal et al. 1976; Caughey and Palmer 1979; Dosio et al. 2005; Zhou et al. 2014).

In this study, a third-degree polynomial of the form is used as an approximation for $\Lambda_{mw}/h$:

$$\frac{\Lambda_{mw}}{h} = 2\pi\beta_P \frac{z}{h}\left\{1 - 0.8\left(\frac{z}{h}\right)\right\}^2 \tag{12}$$

where $\beta_P = 1.2$ is a constant coefficient, $(\pi\beta_P = 3.77)$.

Data obtained in the Minnesota-1973 and Ashchurch-1976 experiments according to (Caughey and Palmer 1979) and in the TRAC98 experiment according to (Bernard-Trottolo et al.

2004), as well as their comparison with the approximation (12), were discussed in detail in (Vulfson and Nikolaev 2022, 2024a). The validity of the approximation (12) can be also demonstrated by comparison with the field measurements $\Lambda_{mw}/h$ at various heights $z/h$, that were conducted by Greenhut and Mastrantonio (1989). The excellent agreement is demonstrated by Fig 3.

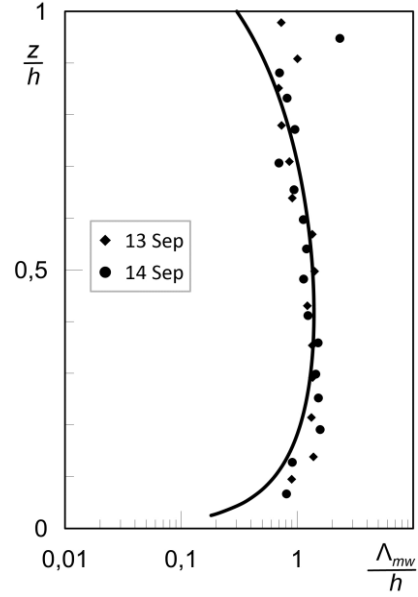

**Fig. 3** The dimensionless value of $\Lambda_{mw}/h$ according to the field measurements. The solid line corresponds to the approximation (12). The black geometric symbols correspond to the selected data from the field measurements by Greenhut and Mastrantonio (1989).

Let us define the "spectral" Prandtl mixing length $l_{PS}$ as a quantity proportional to the integral scale of turbulence $\Lambda_{mw}$. In other words, we will consider that

$$l_{PS} = \gamma_P \Lambda_{mw}, \quad \frac{l_{PS}}{h} = \gamma_P \frac{\Lambda_{mw}}{h}, \quad \gamma_P = \alpha_P/(2\pi\beta_P) \qquad (13)$$

Here, $\alpha_P > 0$ and $\gamma_P > 0$ are positive constants.

We choose the proportionality coefficient $\alpha_P$ so that in the convective surface layer $0 < z/h < 0.1$, the Prandtl theory and the Monin-Obukhov theory lead to the same analytical form of the turbulent exchange coefficient $K_H$. For more details, see (Prandtl 1932; Obukhov 1946). This condition uniquely determines the coefficient $\alpha_P$ such that $\alpha_P = 2k_0$, where $k_0$ is the von Kármán constant. With $k_0 = 0.4$, the coefficient $\gamma_P = k_0/(\pi\beta_P) = 0.127$. Therefore, according to (13), the spectral Prandtl mixing length $l_{PS}$ is approximately 7.8 times smaller than the integral scale of turbulence $\Lambda_{mw}$. Importantly, the magnitude $l_{PS}$ lies in the inertial subrange of the spectrum $\eta \leq l_P \leq \Lambda_{mw}$, where $\eta = \nu^{3/4} / \varepsilon^{1/4}$ is the Kolmogorov scale. For more details, see (Frisch 1995).

The transformation (13) with consideration of (12) leads to the formation of the vertical profile of the "spectral" mixing length

$$\frac{l_{PS}}{h} = \alpha_P \frac{z}{h} \left[ 1 - 0.8 \left( \frac{z}{h} \right) \right]^2, \quad \alpha_P = 2k_0 = 0.8 \quad (14)$$

In the surface layer, close to the underlying surface $z/h \ll 1$, the surface asymptotics (14) take on a linear form. For more details about the limiting form (14), see Appendix A in (Vulfson and Nikolaev 2022).

It is significant that the description of the turbulent heat exchange process in the semi-empirical Prandtl theory is constructed by analogy with the description of the molecular heat exchange process in the elementary kinetic theory of gases, see, for example, (Isihara 1971). In particular, the Prandtl theory assumes that turbulent heat exchange is realized by a stochastic ensemble of isolated eddies moving in a practically stationary environment. In this case, the mixing length corresponds to the average size of the eddy's ensemble. From equation (14), it follows that when $z/h \approx 0.4$, the value is $\max_{0 \leq z/h \leq 1} l_{PS} / h \approx 0.15$. This value is practically identical to the average diameter of floating eddies in the ensemble of small convective thermals, experimentally studied in works by Vulfson N. (1964) and Lenschow and Stephens (1980). Thus, the definition of the spectral mixing length (14) fully corresponds to the physical positions of the Prandtl theory.

It is also necessary to emphasize that the choice of the coefficient $\gamma_P = k_0 / (\pi \beta_P) \approx 0.1$ in our definition of the spectral mixing length $l_{PS} = \gamma_P \Lambda_{mw}$ relies on the MOST and therefore differs from the previously proposed definitions. Thus, for example, in the models (Hanna 1968; Goulart et al. 2004; Degrazia et al. 2015), it was considered that $\gamma_P = 1$. Whereas in the models (Sun 1986, 1989; Zhou et al. 2014), it was considered that $\gamma_P \approx 0.2$.

**3.2. The second moment of vertical velocity as a basic parameter**

Detailed data on the second moment of vertical velocity $\overline{w^2}$ in the convective layer of the atmosphere were obtained in the experiments NCAR-1972, Minnesota-73, AMTEX-1975, METROMEX-1975, Phoenix-1978, TRAC-1998, ARTIST-1999, IHOP_2002, AVEC-2006, and AMMA-2006 and are provided in (Sommeria and LeMone 1978; Kaimal et al. 1976; Lenschow et al. 1980; Hildebrand and Ackerman 1984; Young 1988; Bernard-Trottolo et al. 2004; Gryanik and Hartmann 2002; Kang et al. 2007; Ansmann et al. 2010; Lenschow et al. 2012). The second moment of vertical velocity was also measured in seasonal field experiment by Berg et al. (2017). These full-scale experiments provide almost complete identity of results for the second moments of vertical velocity.

The approximation of the second moment of vertical velocity $\overline{w^2}$ can be performed by the relationship proposed in the work by Lenschow et al. (1980, 2012):

$$\frac{\overline{w^2}}{w_D^2} = \lambda_{ww}^o \left(\frac{z}{h}\right)^{2/3} \left\{1 - 0.8\left(\frac{z}{h}\right)\right\}^2, \quad \lambda_{ww}^o = 1.8 \tag{15}$$

Different approximations of the second moment of vertical velocity $\overline{w^2}/w_D^2$ in the atmosphere and their comparison are extensively discussed in the study (Wood et al. 2010). It was shown that the approximation (15) is statistically the most accurate among existing approximations.

Figure 4 shows measurements of $\overline{w^2}/w_D^2$ in the field experiments IHOP_2002 and AVEC-2006 according to (Kang et al. 2007; Ansmann et al. 2010), as well as in the field experiment by Berg et al. (2017), in which the data underwent seasonal averaging.

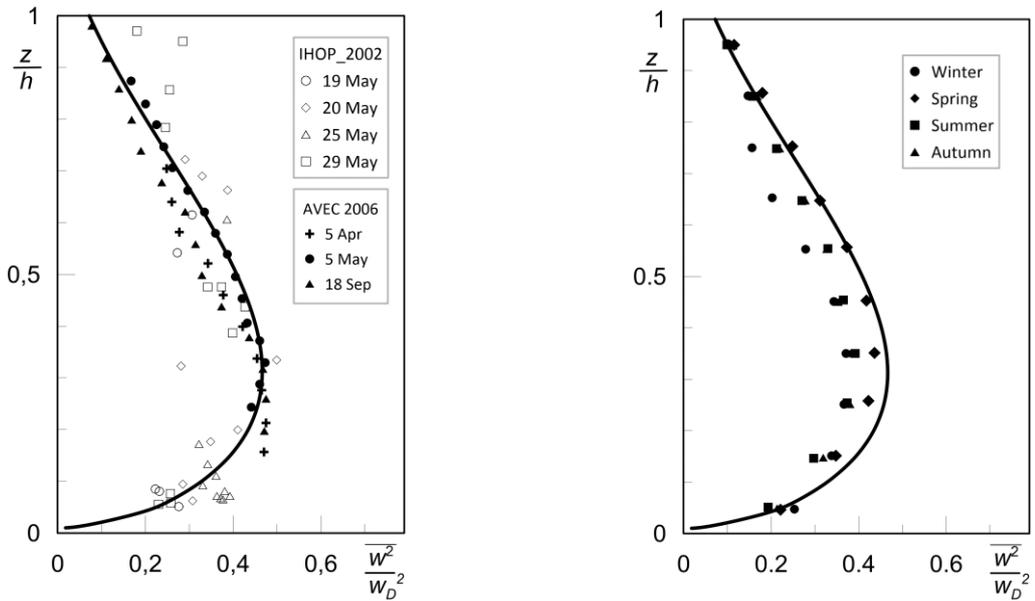

**Fig. 4** The second moment of vertical velocity $\overline{w^2}/w_D^2$ according to field measurements. The solid lines are the approximation (15). Left panel: the white symbols are the IHOP_2002 data presented following (Kang et al. 2007) and the black symbols are the AVEC-2006 data following (Ansmann et al. 2010). Right panel: the symbols are seasonally averaged measurements data by Berg et al. (2017)

According to the results presented in Figure 4, approximation (15) convincingly corresponds to the data from field experiments. Very high agreement is achieved with seasonal averaging, especially in the spring period.

Next, the Prandtl mixing length (14) and the second moment of vertical velocity (15) will be accepted as the basic parameters of the NTLS.

### 3.3. Universal equations for mixed moments involving buoyancy and vertical velocity

The NTLS allows the construction of the two-parameter approximations for mixed turbulent moments of buoyancy and vertical velocity.

Let $g\theta_a$ denote the "adiabatic" buoyancy defined by equation (2), the dimension of which is determined by the relationship $[g\theta_a] = m/s^2$. Then, for approximating the moments of "adiabatic" buoyancy and vertical velocity of the $p$ order, taking into account the Deardorff parameters (5), the following equations hold:

$$\begin{cases} \overline{(g\theta_a)^m w^n} = \bar{\lambda}_{mn}^a \cdot l_{PS}^{-m} \left(\overline{w^2}\right)^{m+n/2}, \quad p = m+n \geq 2, \quad 0 \leq m, n \leq p \\ \dfrac{\overline{(g\theta_a)^m w^n}}{(g\theta_D)^m w_D^n} = \bar{\lambda}_{mn}^a \left(\dfrac{l_{PS}}{h}\right)^{-m} \left(\dfrac{\overline{w^2}}{w_D^2}\right)^{m+n/2} \end{cases} \quad (16)$$

Here, $\bar{\lambda}_{mn}^a > 0$ are unknown constants included in the relationships of turbulent moments of "adiabatic" buoyancy (16); $m \geq 0$ and $n \geq 0$ are non-negative integer indices.

Substituting the basic parameters (14) and (15) into equations (16), we obtain:

$$\dfrac{\overline{(g\theta_a)^m w^n}}{(g\theta_D)^m w_D^n} = \lambda_{mn}^a \left(\dfrac{z}{h}\right)^{-\frac{1}{3}(m-n)} \left[1 - 0.8\left(\dfrac{z}{h}\right)\right]^n, \quad p = m+n \geq 2, \quad 0 \leq m, n \leq p \quad (17)$$

where $\lambda_{mn}^a = \alpha_P^{-m} (\lambda_{ww})^{m+n/2} \bar{\lambda}_{mn}^a$ are unknown constants.

Approximations of the moments of "adiabatic" buoyancy and vertical velocity (16) and (17) were comprehensively compared with the laboratory experimental data in the study (Vulfson and Nikolaev 2022).

The relationships (16)-(17) can be complemented by another set of approximations. Let $g\theta_b$ represent the "Boussinesq" buoyancy, defined by relation (1), whose dimension is determined by the equation $[g\theta_b] = m/s^2$. Then, for the approximation of moments of "Boussinesq" buoyancy and vertical velocity of the $p$-th order, equations similar to (16) and (17) are:

$$\begin{cases} \overline{(g\theta_b)^m w^n} = \bar{\lambda}_{mn}^b \cdot l_{PS}^{-m} \left(\overline{w^2}\right)^{m+n/2}, \quad p = m+n \geq 2, \quad 0 \leq m, n \leq p \\ \dfrac{\overline{(g\theta_b)^m w^n}}{(g\theta_D)^m w_D^n} = \bar{\lambda}_{mn}^b \left(\dfrac{l_{PS}}{h}\right)^{-m} \left(\dfrac{\overline{w^2}}{w_D^2}\right)^{m+n/2} \end{cases} \quad (18)$$

Here, $\bar{\lambda}_{mn}^b > 0$ represents unknown constants included in the relations of turbulent moments of "Boussinesq" buoyancy (18); $m \geq 0$ and $n \geq 0$ are non-negative integers.

Let us transform the moments of "Boussinesq" buoyancy (18) taking into account (14) and (15) then:

$$\frac{\overline{(g\theta_b)^m w^n}}{(g\theta_D)^m w_D^n} = \lambda_{mn}^b \left(\frac{z}{h}\right)^{-\frac{1}{3}(m-n)} \left[1 - 0.8\left(\frac{z}{h}\right)\right]^n, \quad p = m+n \geq 2, \ 0 \leq m,n \leq p \quad (19)$$

where $\lambda_{mn}^b = \alpha_P^{-m}(\lambda_{ww})^{m+n/2} \bar{\lambda}_{mn}^b$ are unknown constants.

Approximations of the mixed moments of "Boussinesq" buoyancy and vertical velocity (18) and (19) were comprehensively compared with the field experimental data in the studies (Vulfson and Nikolaev 2024a,b).

The relations (17) and (19) represent two sets of different turbulent moments, also expressed in terms of parameters $\overline{w^2}$ and $l_{PS}$. The number of these relations depends on the range of the order $p$. For instance, when studying turbulent moments from the second to the fourth order inclusive, the sets of universal equations (17) and (19) each contain 12 relationships.

Comparison with experimental data confirms the validity of the approximations (17) and (19). Thus, the relations (17) and (19) should be considered as a significant theoretical supplement to the semi-empirical Prandtl theory.

With fixed $m$ and $n$, the functional dependencies of (17) and (19) differ by coefficients $\lambda_{mn}^a$ and $\lambda_{mn}^b$. According to the results by Vulfson and Nikolaev (2024a), it follows that $\lambda_{20}^a > \lambda_{20}^b$ and $\lambda_{11}^a = \lambda_{11}^b$. Therefore, it can be assumed that the coefficients of turbulent moments of "adiabatic" and "Boussinesq" buoyancy are related by the inequality $\lambda_{mn}^a \geq \lambda_{mn}^b$. Naturally, the validity of this assumption requires detailed subsequent justification.

## 4. Local similarity and interpretation of the semi-empirical Prandtl turbulence theory

The "spectral" Prandtl mixing length (14) has a quite specific form. Therefore, the question of the correctness of relations of the semi-empirical Prandtl theory, including $l_{PS}$, requires additional justification.

### 4.1. Turbulent heat transfer coefficient

Following the semi-empirical Prandtl theory (Prandtl 1925, 1932), and the NTLS (Vulfson and Nikolaev 2022), we can present the turbulent heat transfer coefficient $K_H$ in the following form:

$$K_H = l_{PS}\left(\overline{w^2}\right)^{1/2}, \quad \frac{K_H}{h \cdot w_D} = \frac{l_{PS}}{h} \cdot \frac{\sqrt{\overline{w^2}}}{w_D} \tag{20}$$

Here, $l_{PS}$ represents the "spectral" mixing length, and $\overline{w^2}$ is the second moment of vertical velocity.

The transformation (20) with consideration of (14) and (15) leads to the approximation:

$$\frac{K_H}{h \cdot w_D} = \lambda_K^o \left(\frac{z}{h}\right)^{4/3} \left\{1 - 0.8\left(\frac{z}{h}\right)\right\}^3, \quad \lambda_K^o = 2k_0\sqrt{\lambda_{ww}^o} \tag{21}$$

where $\lambda_K^o$ is a constant coefficient. When $\lambda_{ww}^o = 1.8$ and $k_0 = 0.4$, it follows that $\lambda_K^o = 2k_0 \cdot (\lambda_{ww}^o)^{1/2} = 1.07$.

It can be shown that in the surface layer, the approximation (21) is identical to the coefficient of turbulent heat transfer of the MOST under conditions of free convection. For more details, refer to (Obukhov 1946; Monin and Yaglom 1975; Panofsky 1978; Vulfson and Nikolaev 2022, 2024a).

The correctness of the local similarity approximation (21) can be validated by comparing it with the results of numerical integration of large-eddy simulation models. Comparison of the approximation (21) with the results of numerical models (Holtslag and Moeng 1991; Abdella and Mcfarlane 1997; Noh et al. 2003; LeMone et al. 2013; Burchard and Petersen 1999) is carried out in works (Vulfson and Nikolaev 2022, 2024a,b). It was demonstrated that the computational results are in good agreement with the approximation (21). Importantly, as the presented numerical calculations are based on various semi-empirical turbulence models and a priori relationships not related to the theory of local similarity, therefore the results of such comparison are more convincing.

**4.2. Dissipation of Kinetic Energy**

The equation for the change in kinetic energy of turbulent motion in a stratified atmosphere follows from system (6), supplemented with terms for molecular viscosity and thermal conductivity. The general form of the equation for the change in kinetic energy of turbulent motion is given by:

$$\begin{cases} \dfrac{\partial}{\partial t} b = -\dfrac{\partial}{\partial z}\left\{\overline{w(\Phi_b + b)}\right\} + \overline{g\theta_b w} - \varepsilon_b \\ \varepsilon_b = -\nu \cdot \overline{\left(\mathbf{u'} \cdot \nabla^2 \mathbf{u'} + w\nabla^2 w\right)} \end{cases} \quad (22)$$

Here, $b = \left\{\overline{(\mathbf{u'} \cdot \mathbf{u'})} + \overline{w^2}\right\}/2$ represents the kinetic energy of turbulent motion, including velocity fluctuations in both horizontal $\mathbf{u'}$ and vertical $w$ directions, $\nu$ is the kinematic viscosity coefficient, $\nabla^2$ is the Laplace operator, and $\varepsilon_b$ denotes the local parameter of kinetic energy dissipation, with dimensions of $[\varepsilon_b] = \mathrm{m}^2/\mathrm{s}^3$.

An equation similar to (22) was first proposed in works (Kolmogorov 1942; Prandtl 1945) to describe turbulent motion in homogeneous fluids. The use of (22) to describe turbulent motion in a stratified atmosphere (6) was first carried out in the work (Obukhov 1946), see also (Calder 1949; Ogura 1952).

The semi-empirical theory of turbulence also widely employs a simplified form of (22), including an approximation of kinetic energy dissipation.

$$\varepsilon_b = \overline{\lambda}_{\varepsilon b} l_P^{-1} \left(\overline{w^2}\right)^{3/2} = \overline{\lambda}_{\varepsilon b} \left\{l_P^{-1}\left(\overline{w^2}\right)^{1/2}\right\} \cdot \overline{w^2} \quad (23)$$

Here, $\overline{\lambda}_{\varepsilon b} > 0$ is a positive constant corresponding to the arbitrary form of the mixing length $l_P$. The dissipation approximation (23) was proposed by Obukhov (1946), see also (Calder 1949; Ogura 1952).

Let us consider the approximation $\varepsilon_b$ based on NTLS. Note that $[\varepsilon_b] = \mathrm{m}^2/\mathrm{s}^3$, as well as $\sqrt{\overline{w^2}}$ and $l_{PS}$ are the basic parameters, then:

$$\varepsilon_b = \lambda_{\varepsilon b} \cdot l_{PS}^{-1}\left(\overline{w^2}\right)^{3/2}, \quad \frac{\varepsilon_b}{g\theta_D w_D} = \frac{\lambda_{\varepsilon b}}{g\theta_D h} \frac{h}{l_{PS}} \frac{\left(\overline{w^2}\right)^{3/2}}{w_D} = \lambda_{\varepsilon b}\left(\frac{l_{PS}}{h}\right)^{-1}\left(\frac{\overline{w^2}}{w_D^2}\right)^{3/2} \quad (24)$$

Here, $\lambda_{\varepsilon b} > 0$ is an unknown constant included in the expression for the dissipation of kinetic energy, normalized by the basic parameters.

Let us transform the second equality (24) taking into account the vertical form of the basic parameters (14) and (15), as well as the Deardorff scaling (5), then:

$$\frac{\varepsilon_b}{g\theta_D w_D} = \lambda_{\varepsilon b}^o \left[1 - 0.8\left(\frac{z}{h}\right)\right], \quad \lambda_{\varepsilon b}^o = \alpha_P^{-1}(\lambda_{ww}^o)^{3/2}\lambda_{\varepsilon b} \quad (25)$$

Here, $\alpha_P = 2k_0$ is the doubled constant of von Karman, and $\lambda_{\varepsilon b}^o$ is a constant included in the transformed expression for the dissipation of kinetic energy (25). From relation (25), the surface asymptote $\varepsilon_b \approx \lambda_{\varepsilon b}^o g \theta_D w_D = \lambda_{\varepsilon b}^o g S_\theta$ follows.

Comprehensive data on the dissipation of kinetic energy in the convective layer of the atmosphere were obtained in experiments Minnesota-1973, Ashchurch-1976, Limagne-1974, Beauce-1976, AMTEX-1975, Toulouse-1976, BAO-1981, as well as in field experiments LIFT-1996, TRAC 98, ARTIST-1999, ShUREX-2016, and are presented in the works (Kaimal et al. 1976; Caughey and Palmer 1979; Sorbjan 1988, 1991; Lenschow et al. 1980; Bernard-Trottolo et al. 2004; Druilhet et al. 1983; Luce et al. 2020).

Comparison of the local similarity approximation (25) with the coefficient $\lambda_{\varepsilon b}^o = 1$ and the data from field measurements is presented on Figure 5. The left panel displays the data from field experiments Minnesota-1973, Ashchurch-1976, and ShUREX-2016 according to (Caughey and Palmer 1979; Luce et al. 2020). The right panel shows the data from the field experiment Toulouse-1976 according to (Druilhet et al. 1983).

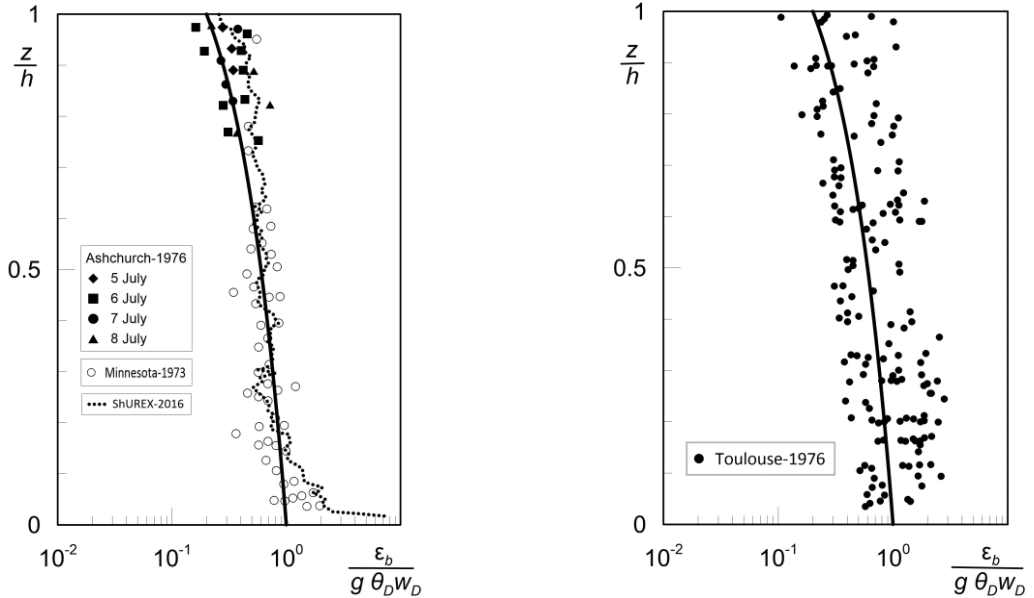

**Fig. 5** Dimensionless dissipation of kinetic energy according to field measurements. The solid lines are the approximation (25) with $\lambda_\varepsilon^o = 1$. Left panel: the black geometric symbols are the data of the Ashchurch-1976 experiment and the white circles are the data of the Minnesota-1973 experiments according to (Caughey and Palmer 1979). The dotted line is the results of the ShUREX-2016 experiments based on (Luce et al. 2020). The right panel: the black dots are the results of the Toulouse-1976 experiment following (Druilhet et al. 1983)

The comparison results presented in Figure 5 demonstrate that the approximation (25) of the NTLS with the coefficient $\lambda_{\varepsilon b}^o = 1$ convincingly corresponds to the field measurements of kinetic energy dissipation $\varepsilon_b$ throughout the entire convective boundary layer $0 \leq z/h \leq 1$. It is noteworthy that according to (25), when $\lambda_{ww}^o = 1.8$ and $\alpha_P = 2k_0 = 0.8$ are constants, the coefficient $\lambda_{\varepsilon b} = \alpha_P (\lambda_{ww}^o)^{-3/2} \lambda_{\varepsilon b}^o = 0.33$.

### 4.3. Dissipation of Buoyancy Square

For the square of "Boussinesq" buoyancy $g\theta_b$, there exists an evolutionary equation similar to (25). This equation follows from system (6), supplemented with terms for molecular viscosity and thermal conductivity, and is given by:

$$\begin{cases} \dfrac{\partial}{\partial t}\dfrac{1}{2}\overline{(g\theta_b)^2} = -\dfrac{\partial}{\partial z}\dfrac{1}{2}\overline{(g\theta_b)^2 w} - g\Gamma(z)\dfrac{\partial}{\partial z}\overline{g\theta_a} - \varepsilon_{g\theta}, \\ \varepsilon_{g\theta} = -\chi \cdot \overline{g\theta_b \cdot \nabla^2 g\theta_b} \end{cases} \quad (26)$$

Here, $g\Gamma(z) = \partial \overline{g\theta_a}/\partial z$ is the stratification parameter, identical to the adiabatic buoyancy gradient; $\varepsilon_{g\theta}$ is the dissipation of the mean "Boussinesq" buoyancy square, with dimensions of $[\varepsilon_{g\theta}] = m^2/s^5$; $\chi$ is the coefficient of molecular thermal conductivity; $\nabla^2$ is the Laplace operator.

Equation (26) which includes the dissipation of the square of "Boussinesq" buoyancy, was presented in the work (McColl et al. 2017).

It is evident that in a neutral atmosphere, the equation for the square of "adiabatic" buoyancy $g\theta_a$ follows from (26) when $g\Gamma(z) = 0$ and corresponds to system (9). A similar equation for turbulent temperature fluctuations was first used in studies (Obukhov 1949, 1959).

The semi-empirical theory of turbulence also widely employs a simplified form of (26), including an approximation of the dissipation of the square of "Boussinesq" buoyancy:

$$\varepsilon_{g\theta} = \overline{\lambda}_{\varepsilon g\theta} \left\{ l_P^{-1} \left(\overline{w^2}\right)^{1/2} \right\} \cdot \overline{(g\theta_b)^2} \quad (27)$$

Here, $\overline{\lambda}_{\varepsilon b}$ is a positive constant corresponding to the arbitrary form of the mixing length. The approximation of the dissipation of the buoyancy square (27) in a form similar to (23) was proposed in the scheme (Mellor 1973). This scheme has been adapted for use in numerical modelling schemes, see, for example, (Abdella and Mcfarlane 1997; Ferrero et al. 2009).

Let us consider the approximation $\varepsilon_{g\theta}$ based on NTLS. Given that $[\varepsilon_{g\theta}] = m^2/s^5$, we have:

$$\varepsilon_{g\theta} = \lambda_{\varepsilon g\theta} l_{PS}^{-3} \left(\overline{w^2}\right)^{5/2}, \quad \dfrac{\varepsilon_{g\theta} h}{(g\theta_D)^2 w_D} \cdot \dfrac{h^3}{w_D^5} = \lambda_{\varepsilon g\theta} \left(\dfrac{l_{PS}}{h}\right)^{-3} \left\{ \dfrac{\left(\overline{w^2}\right)^{1/2}}{w_D} \right\}^5 \quad (28)$$

Here, $\lambda_{\varepsilon g\theta} > 0$ is a constant included in the dissipation of the buoyancy square of local similarity $\varepsilon_{g\theta}$, normalized by the basic parameters.

Note that when $l_P = l_{PS}$, the transformation (27) considering the approximation (16) in the $\overline{(g\theta_b)^2} \sim l_{PS}^{-2} \left(\overline{w^2}\right)^2$ leads to the relation (28). Thus, the approximations (27) and (28) are equivalent.

Let us transform the second equality (28) taking into account the vertical forms of the basic parameters (14) and (15), as well as the Deardorff scaling (5), then:

$$\frac{\varepsilon_{g\theta} h}{(g\theta_D)^2 w_D} = \lambda_{\varepsilon g\theta}^o \left(\frac{z}{h}\right)^{-4/3} \left[1 - 0.8\left(\frac{z}{h}\right)\right]^{-1}, \quad \lambda_{\varepsilon g\theta}^o = \alpha_P^{-3} (\lambda_{ww}^o)^{5/2} \lambda_{\varepsilon g\theta} \qquad (29)$$

Here, $\lambda_{\varepsilon g\theta}^o$ is a constant included in the transformed form of the buoyancy square dissipation. In the surface layer $0 \leq z/h < 0.1$, from relation (29), it follows the approximate equality $\varepsilon_{g\theta} h / (g\theta_D)^2 w_D \approx \lambda_{\varepsilon g\theta}^o (z/h)^{-4/3}$.

Comprehensive data on the dissipation of the buoyancy square in the convective layer of the atmosphere were obtained in experiments Minnesota-1973, Ashchurch-1976, Limagne-1974, Beauce-1976, and are presented in studies (Kaimal et al. 1976; Caughey and Palmer 1979; Sorbjan 1988, 1991).

Comparison of the local similarity approximation (29) with the coefficient $\lambda_{g\theta}^o = 0.5$ and the data from field measurements is presented on Figure 6. Left panel shows the data from the Minnesota-1973 and Ashchurch-1976 experiments according to (Caughey and Palmer 1979). Right panel presents the data from the Limagne-1974 and Beauce-1976 experiments, according to (Guillemet et al. 1983).

The comparison results presented in Figures 6 demonstrate that the approximation of the local similarity theory (29) with the coefficient $\lambda_{\varepsilon g\theta}^o = 0.5$ convincingly matches the field measurements of $\varepsilon_{g\theta}$ throughout the entire convective boundary layer $0 \leq z/h \leq 1$. It is noteworthy that according to (29) if $\alpha_P = 2k_0 = 0.8$ and $\lambda_{ww}^o = 1.8$, the coefficient $\lambda_{\varepsilon g\theta} = \alpha_P^3 (\lambda_{ww}^o)^{-5/2} \lambda_{\varepsilon g\theta}^o = 0.06$.

The results presented in sections 4.2 and 4.3 indicate the high effectiveness of the new local similarity theory for approximating the dissipation of kinetic energy and the square of " Boussinesq " buoyancy. It should also be emphasized that the approximations of well-known local similarity theories (Zeman and Lumley 1976; Sorbjan 1988, 1991; Vulfson et al. 2004) reasonably approximate the field measurements of $\varepsilon_b$ and $\varepsilon_{g\theta}$ only in the lower half of the convective boundary layer, $0 \leq z/h \leq 0.55$. Therefore, in this situation, the use of the approximation of the new local similarity theory (25) and (29) is more appropriate.

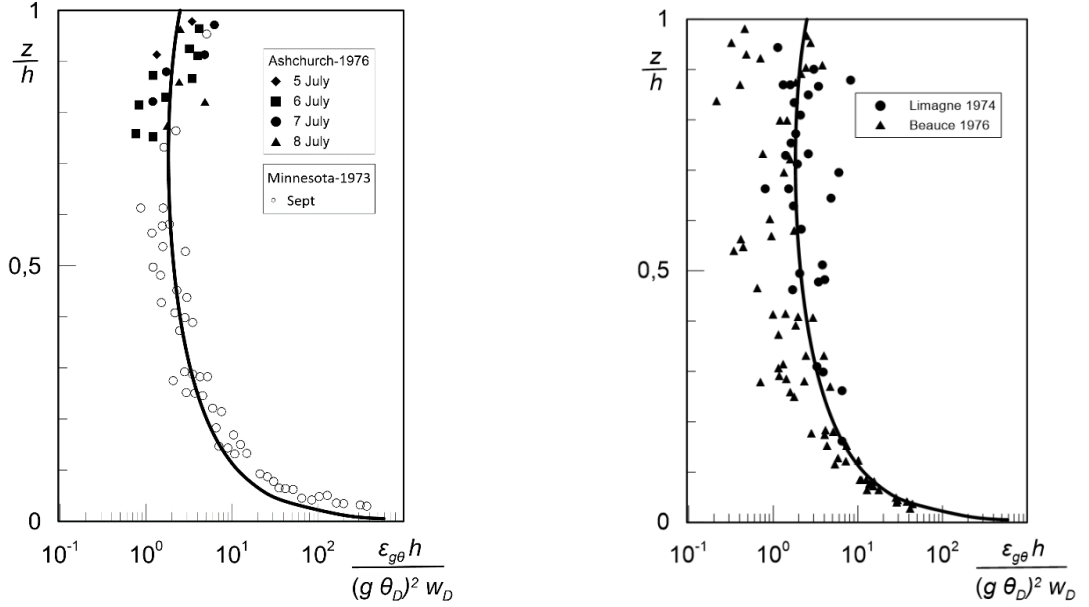

**Fig. 6** Dimensionless dissipation of 'Boussinesq' buoyancy square. The solid lines are the approximation (29) with $\lambda_{\varepsilon g\theta}^o = 0.5$. Left panel: the black symbols and the white circles are the data from the Ashchurch-1976 and Minnesota-1973 experiments based on (Caughey and Palmer 1979). Right panel: the circles and the triangles are the data from Limagne-1974 and Beauce-1976 experiments following (Guillemet et al. 1983)

## 5. Local similarity and interpretation of Hanna's semi-empirical turbulence theory

Let us demonstrate that NTLS is consistent with the semi-empirical turbulence theory by Hanna (1968, 1984).

Within this semi-empirical theory, the "spectral" coefficient of turbulent heat transfer $K_H$ and the dissipation of turbulent kinetic energy $\varepsilon_b$ are defined by the following relationships:

$$K_H = \gamma_H \cdot \Lambda_{mw} \sqrt{\overline{w^2}}, \quad \varepsilon_b = a_H^{-1} \Lambda_{mw}^{-1} \left(\overline{w^2}\right)^{3/2} \tag{30}$$

Here, $\gamma_H \simeq 0.13$ and $a_H \simeq 0.3$ are the experimental constants of Hanna's theory.

For the atmospheric surface layer $0 \leq z/h \leq 0.1$, the relationships (30) were first empirically established in the papers (Hanna 1968, 1984). Therefore, we will refer to the parameter $a_H$ as the constant of Hanna. Subsequent studies have shown that the relationships (30) hold true for most of the convective layer of the atmosphere $0 \leq z/h \leq 0.7$.

According to the measurements (Hanna 1968), the constant $a_H = 0.29$. According to measurements in the CBL conducted by Kaimal and Haugen (1967), Kaimal (1973) and Wamser et al. (1977), the constant is $a_H = 0.32$.

### 5.1. "Spectral" turbulent eddy diffusion coefficient

Let us demonstrate that the turbulence exchange coefficient of Hanna's theory, specified by the first relationship (30), corresponds to the NTLS. To do this, we consider the NTLS relationships for the exchange coefficient (20) and the "spectral" mixing length (13). Transforming (20) with (13) leads to the relationship:

$$K_H = \gamma_P \Lambda_{mw} \left(\overline{w^2}\right)^{1/2} \tag{31}$$

where $\gamma_P = k_0 / (\pi \beta_P) = 0.127$. The new form of local similarity of $K_H$ contains the size of turbulent eddies with the highest kinetic energy of vertical motion $\Lambda_{mw}$, instead of the spectral mixing length $l_{PS}$. Comparing (31) and (30) with the equality $\gamma_H = \gamma_P$ indicates the identity of the similarity relationship for $K_H$ and Hanna's approximation.

The relationship (31) has been used in the studies (Hanna 1968, 1984), as well as in the application to advection-diffusion equations for air pollution models, see (Degrazia et al. 2001; Goulart et al. 2004).

Numerical modelling shows that at large distances from the source of pollution, the coefficient of vertical eddy diffusion (31) takes on an asymptotic form similar to (20), see (Degrazia et al. 2015).

The correspondence of approximation (31) to the results of numerical modelling not only convincingly confirms the validity of the theory of local similarity for turbulent convection problems but also indicates its consistency with the semi-empirical theory (Hanna 1968, 1984).

### 5.2. Turbulence Dissipation Scale and the Hanna Constant

Following the ideas of (Kolmogorov 1942; Hinze 1975), let us define the turbulence dissipation scale $\Lambda_\varepsilon$ such that

$$\Lambda_\varepsilon = \frac{\left(\overline{w^2}\right)^{3/2}}{\varepsilon_b}, \quad \frac{\Lambda_\varepsilon}{h} = \left(\frac{\overline{w^2}}{w_D^2}\right)^{3/2} \frac{w_D^3}{\varepsilon_b h} \tag{32}$$

Within the framework of the Hanna theory (30), it is true

$$a_H = \frac{\Lambda_\varepsilon}{\Lambda_{mw}} \tag{33}$$

Equation (33) allows us to consider the Hanna constant $a_H$ as the ratio of the dissipation scale $\Lambda_\varepsilon$ to the integral scale of turbulence $\Lambda_{mw}$. The value of $a_H = 0.32$ according to measurements in the CBL, conducted by (Kaimal 1973; Wamser et al. 1977).

It is important to note that the experimental determination of $\Lambda_\varepsilon$ in Hanna's theory and in NTLS is implemented in different ways. Therefore, the ratio of the dissipation scale $\Lambda_\varepsilon$ to the integral scale of turbulence $\Lambda_{mw}$ in the NTLS theory and Hanna's theory may differ.

Within the framework of NTLS, the second relation (32) can be transformed considering (15) and (25), then

$$\frac{\Lambda_\varepsilon}{h} = (\lambda_{\varepsilon b}^o)^{-1}(\lambda_{ww}^o)^{3/2} \cdot \left(\frac{z}{h}\right)\left\{1 - 0.8\left(\frac{z}{h}\right)\right\}^2 \tag{34}$$

Let the constant $a_P$ represent the ratio of the dissipation scale $\Lambda_\varepsilon$ to the integral scale of turbulence $\Lambda_{mw}$ within NTLS. Then from (34) and (12), it follows that:

$$a_P = \frac{\Lambda_\varepsilon}{\Lambda_{mw}} = \frac{(\lambda_{\varepsilon b}^o)^{-1}(\lambda_{ww}^o)^{3/2}}{2\pi\beta_P} \tag{35}$$

The second part of (35) allows us to compute $a_P$ from indirect measurements in the surface layer. When $\lambda_{\varepsilon b}^o = 1$, $\lambda_{ww}^o = 1.8$ and $\pi\beta_P = 3.77$ the value of the constant $a_P = 0.32$. This value is identical to the value of the Hanna constant $a_H$ obtained from direct measurements in the studies (Kaimal 1973; Wamser et al. 1977).

The results of Section 5.2 show that the semi-empirical turbulence theory of Hanna can be interpreted as a consequence of NTLS.

## 6. Local similarity and interpretation of the semi-empirical Richardson-Onsager turbulence theory

Let us consider the semi-empirical turbulence theory by Richardson (1926), see also (Roberts and Webster 2002). According to it, the relationship for the turbulent exchange coefficient is given by $K_H = c_R \cdot l_p^{4/3}$, where $c_R \sim 0.4 \, cm^{2/3}/s$ is the Richardson dimensional coefficient. According to measurements in the open ocean $c_R = 0.1 - 0.3 \, cm^{2/3}/s$, see (Fischer 1979).

Supplementations and justifications of the Richardson relationship were proposed in the studies (Obukhov 1949; Onsager 1949), see also (Batchelor 1950).

According to (Obukhov 1949; Onsager 1949), the dimensional coefficient should be given in the form $c_R \simeq \varepsilon_b^{1/3}$, where $\varepsilon_b$ is the dissipation of turbulent kinetic energy. In this case, the Prandtl mixing length should be located in the inertial interval of the spectrum $\eta \leq l_P \leq \Lambda_{mw}$, where $\eta = \nu^{3/4} / \varepsilon^{1/4}$ is the Kolmogorov scale, see more details in (Frisch 1995).

Within the framework of NTLS, the turbulence exchange coefficient $K_H$ and the dissipation of turbulent kinetic energy $\varepsilon_b$ satisfy the relations (20) and (24), i.e. $K_H = l_{PS} \left(w^2\right)^{1/2}$ $\varepsilon_b = \lambda_\varepsilon \cdot l_{PS}^{-1} \left(w^2\right)^{3/2}$. Hence, it immediately follows that

$$K_H = (\lambda_\varepsilon^{-1/3} \varepsilon_b^{1/3}) \cdot l_{PS}^{4/3} \qquad (36)$$

Notice that according to (13), the "spectral" mixing length is $l_{PS} = \gamma_P \Lambda_{mw}$, where $\gamma_P = 0.127$. Therefore, the spectral mixing length $l_{PS}$ belongs to the inertial subrange. Thus, NTLS is consistent with Richardson's semi-empirical theory.

## 7. Structural Parameters

Structural parameters of velocity and buoyancy play a significant role in studying the propagation of acoustic and electromagnetic waves in the atmosphere (Tatarski et al. 1961; Rytov et al. 1989) Let us show how the theory of local similarity can be used to propose new approximations for second and third-order structural parameters.

### 7.1. Second-order Structural Parameters. Classical Similarity Theory Approximations

The determination of structural parameters can be performed either based on structural functions, see, for example, (Obukhov 1949; Monin 1958; Kolmogorov 1962, 1968) or based on one-dimensional spectral decompositions, see, for example, (Wyngaard et al. 1971; Gibbs and Fedorovich 2020). It is considered that both of these approaches are equivalent, however, some authors do not share such a point of view, see more in (George 2013). Hereafter, we will rely on the structural approach.

Let us fix the level $z$ and choose the horizontal axis $x$ along the wind direction. Let $U$ and $u'$ be the mean wind speed and the horizontal velocity fluctuation along the $x$ axis; $g\theta_b$ be the 'Boussinesq' buoyancy.

Under conditions of homogeneous and locally isotropic turbulence and considering Kolmogorov's hypotheses, see, for example, (Wyngaard and Clifford 1978), we define the second-order structural functions. Then,

$$\begin{cases} D_{uu}(r) = \overline{[\Delta u(r)]^2}, & \Delta u(r) = u'(x) - u'(x+r) \\ D_{\theta\theta}(r) = \overline{[\Delta g\theta_b(r)]^2}, & \Delta g\theta_b(r) = g\theta_b(x) - g\theta_b(x+r) \end{cases} \quad (37)$$

Here, $r$ is the distance along the $x$-axis; $\Delta u(r)$ and $\Delta g\theta_b(r)$ are the differences in horizontal velocity and "Boussinesq" buoyancy fluctuations at two points along the $x$-axis at the same time $t$; $D_{uu}(r) > 0$ and $D_{\theta\theta}(r) > 0$ are the second-order structural functions; the bar symbol denotes averaging.

Let $\eta = \nu^{3/4}\varepsilon_b^{-1/4}$ be the Kolmogorov scale; $\Lambda_{mw}$ be the integral scale of turbulence, see Section 3.1. Following (Wyngaard and Clifford 1978), let us fix the level $z$, and in the inertial subrange $\eta \ll r \ll \Lambda_{mw}$, we define the structural parameters $C_{uu}$ and $C_{\theta\theta}$ such that:

$$\begin{cases} D_{uu}(r) = C_{uu} r^{2/3} \\ D_{\theta\theta}(r) = C_{\theta\theta} r^{2/3} \end{cases} \quad (38)$$

where $C_{uu} > 0$ and $C_{\theta\theta} > 0$ are dimensional structural parameters independent of $r$.

From equations (38), it follows that the dimensionalities of the structural parameters for velocity and buoyancy are $[C_{uu}] = m^{4/3} s^{-2}$ and $[C_{\theta\theta}] = m^{4/3} s^{-4}$. For more details on the observed structural parameters of the convective layer and the methods of their measurement, refer to (Kaimal and Finnigan 1994; Weiss 2002; Beyrich et al. 2012).

Let us consider the approximations of the structural parameters within the framework of classical similarity theory. Under conditions of free or forced convection with weak wind, the following relationships can be assumed to be hold:

$$\begin{cases} C_{uu} = q_{uu}(gS_\theta, z, h) \\ C_{\theta\theta} = q_{\theta\theta}(gS_\theta, z, h) \end{cases} \quad (39)$$

where $q_{uu} > 0$ and $q_{\theta\theta} > 0$ are undetermined dimensional positive functions.

Equations (39) using Buckingham's theorem can be transformed:

$$\begin{cases} C_{uu} = \nu_{uu}^o gS_\theta Q_{uu}(z/h), & \lim_{z/h \to 0} Q_{uu}(z/h) = 1 \\ C_{\theta\theta} = \nu_{\theta\theta}^o (gS_\theta)^{4/3} z^{-4/3} Q_{\theta\theta}(z/h), & \lim_{z/h \to 0} Q_{\theta\theta}(z/h) = 1 \end{cases} \quad (40)$$

Here $\nu_{uu}^o > 0$ and $\nu_{\theta\theta}^o > 0$ are undetermined positive constants; $Q_{uu}(z/h) > 0$ and $Q_{\theta\theta}(z/h) > 0$ are undetermined positive functions. The second relationship (40) was first obtained in the study (Frisch and Ochs 1975). In this work, a quadratic function of $z/h$ was used for $Q_{\theta\theta}(z/h)$. Moreover, the dimensional power dependence $z^{-4/3}$ for the structural parameter $C_{\theta\theta}$ in the lower half of the CBL was first experimentally established in the study (Tsvang 1969).

The relationship (40) can be conveniently transformed considering Deardorff parameters (5), then

$$\begin{cases} \dfrac{C_{uu}h^{2/3}}{w_D^2} = \nu_{uu}^o Q_{uu}(z/h) & \lim_{z/h \to 0} Q_{uu}(z/h) = 1 \\ \dfrac{C_{\theta\theta}h^{2/3}}{(g\theta_D)^2} = \nu_{\theta\theta}^o \left(\dfrac{z}{h}\right)^{-\dfrac{4}{3}} Q_{\theta\theta}(z/h) & \lim_{z/h \to 0} Q_{\theta\theta}(z/h) = 1 \end{cases} \quad (41)$$

It is essential to note that classical similarity theory does not allow determining the functions $Q_{uu}(z/h)$ and $Q_{\theta\theta}(z/h)$, therefore assumes their a priori specification. The form of structural coefficients can be defined either from known observational data, see (Frisch and Ochs 1975; Weill et al. 1980), or numerical experiments, see (Gibbs and Fedorovich 2014; Gibbs et al. 2016).

According to (41), in the convective surface layer $0 \leq z/h \leq 0.1$, classical similarity theory approximations for structural parameters take the form

$$\dfrac{C_{uu}h^{2/3}}{w_D^2} = \nu_{uu}^o, \qquad \dfrac{C_{\theta\theta}h^{2/3}}{(g\theta_D)^2} = \nu_{\theta\theta}^o \left(\dfrac{z}{h}\right)^{-4/3} \quad (42)$$

The dimensionless forms (42) with coefficients $\nu_{uu}^o = 1.5$ and $\nu_{\theta\theta}^o = 2.7$ were first presented in (Kaimal et al. 1976). Measurements of structural parameters $C_{uu}$ and $C_{\theta\theta}$ were performed in field experiments, Minnesota-73; Ashchurch-76 and USG-88 and are provided in the papers (Kaimal et al. 1976; Caughey and Palmer 1979; Shao et al. 1991) respectively. Measurements of the structural parameter $C_{\theta\theta}$ were performed in field experiments, Flip-72 (San Diego), GATE-74, AMTEX-1975, Toulouse-1976 and ShUREX-2016 and are provided in the papers (Wyngaard and LeMone 1980; Lenschow et al. 1980; Druilhet et al. 1983; Luce et al. 2020).

Figure 7 shows the values of the structural parameters $C_{uu}$ and $C_{\theta\theta}$ according to the data from experiments Minnesota-1973 and Ashchurch-1976 as presented in (Caughey and Palmer 1979). The approximations (42) are also presented in that figure with $\nu_{uu}^o = 1.5$ and $\nu_{\theta\theta}^o = 2.7$.

Acceptable agreement with observational data allows extending the approximation (42) from the surface layer $0 \leq z/h \leq 0.1$ to its lower half $0 \leq z/h \leq 0.6$, see Fig. 7. Therefore, relationships (42) can be considered as simple approximations of classical similarity theory. However, such an approach is not sufficiently theoretically justified.

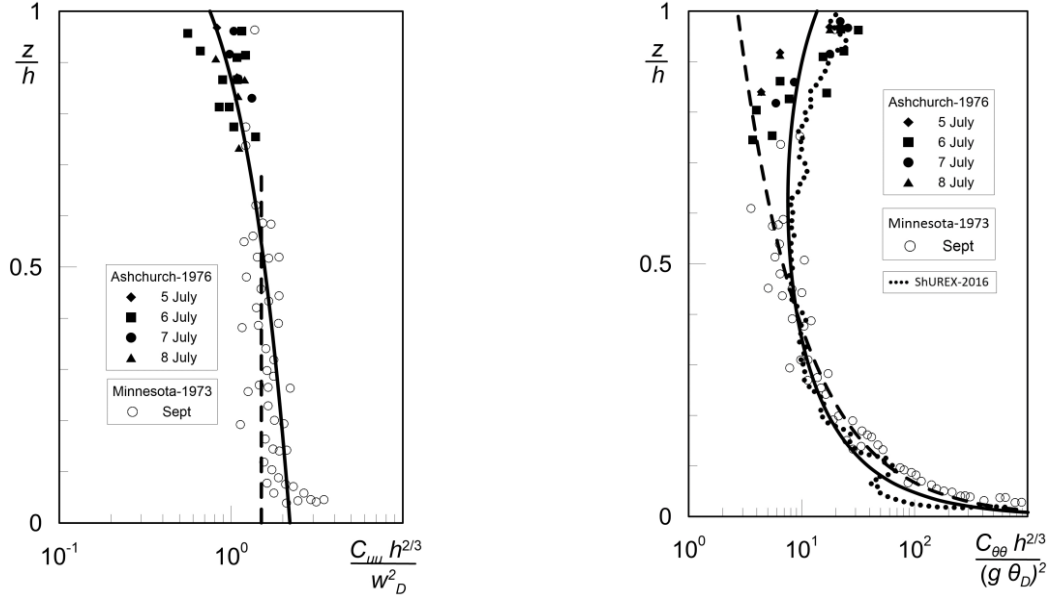

**Fig. 7** The vertical profiles of normalized structural parameters $C_{uu}$ and $C_{\theta\theta}$. Left panel: the white and black symbols are the experimental data by Caughey and Palmer (1979). The dashed line is the classical similarly approximation (42) with $v^o_{uu} = 1.5$. The solid line is the local similarity approximation (44) with $v^o_{uu} = 2.2$. Right Panel: the white and black symbols are the experimental data by Caughey and Palmer (1979). The dotted line is the experimental results by Luce et al. (2020). The dashed line is the classical similarly approximation (42) with $v^o_{\theta\theta} = 2.7$. The solid line is the local similarity approximation (44) with $v^o_{\theta\theta} = 1.6$.

### 7.2. Second-order structural parameters. Approximations of the local similarity theory

Now let us consider the approximations of structural parameters within the framework of the local similarity theory proposed by Vulfson and Nikolaev (2022). This theory allows determining the analytical form of structural coefficients and is therefore more efficient than the classical similarity theory.

According to the NTLS, the spectral mixing length $l_{PS}$ and the second moment of vertical velocity $\overline{w^2}$, with dimensions $[l_{PS}] = m$ and $\left[\overline{w^2}\right] = m^2/s^2$ respectively, are chosen as basic parameters. According to (38), the dimensions of the structural parameters of velocity and buoyancy take the form $[C_{uu}] = m^{4/3}s^{-2}$ and $[C_{\theta\theta}] = m^{4/3}s^{-4}$. Then, relying on the dimension theory using the Deardorff scaling parameters (5), we obtain

$$\begin{cases} C_{uu} = \hat{v}_{uu} l_{PS}^{-2/3} \overline{w^2}, & \dfrac{C_{uu} h^{2/3}}{w_D^2} = \hat{v}_{uu} \left(\dfrac{l_{PS}}{h}\right)^{-2/3} \dfrac{\overline{w^2}}{w_D^2} \\ C_{\theta\theta} = \hat{v}_{\theta\theta} l_{PS}^{-8/3} \left(\overline{w^2}\right)^2, & \dfrac{C_{\theta\theta} h^{2/3}}{(g\theta_D)^2} = \dfrac{C_{\theta\theta} h^{8/3}}{w_D^4} = \hat{v}_{\theta\theta} \left(\dfrac{l_{PS}}{h}\right)^{-8/3} \dfrac{\left(\overline{w^2}\right)^2}{w_D^4} \end{cases} \quad (43)$$

Here, $\hat{v}_{uu}$ and $\hat{v}_{\theta\theta}$ are undetermined coefficients included in the forms of structural parameters normalized by the base parameters.

The relationships of local similarity (43) can be transformed using the vertical forms of basic parameters (14) and (15), giving:

$$\begin{cases} \dfrac{C_{uu} h^{2/3}}{w_D^2} = \hat{v}_{uu} \left(\dfrac{l_{PS}}{h}\right)^{-2/3} \dfrac{\overline{w^2}}{w_D^2} = v_{uu}^o \left\{1 - 0.8\left(\dfrac{z}{h}\right)\right\}^{2/3} \\ \dfrac{C_{\theta\theta} h^{2/3}}{(g\theta_D)^2} = \hat{v}_{\theta\theta} \left(\dfrac{l_{PS}}{h}\right)^{-8/3} \dfrac{\left(\overline{w^2}\right)^2}{w_D^4} = v_{\theta\theta}^o \left(\dfrac{z}{h}\right)^{-4/3} \left\{1 - 0.8\left(\dfrac{z}{h}\right)\right\}^{-4/3} \end{cases} \quad (44)$$

Here, $v_{uu}^o = \hat{v}_{uu} \alpha_P^{-2/3} \lambda_{ww}^o$ and $v_{\theta\theta}^o = v_{\theta\theta}^b \alpha_P^{-8/3} (\lambda_{ww}^o)^2$ are undetermined constants included in the transformed form of structural parameters.

An a priori approximation $C_{\theta\theta}$, that is similar to the second relationship (44), but including the power exponent $-2/3$ in curly brackets, was proposed by Weill et al. (1980). An approximation of local similarity for the structural parameter $C_{\theta\theta}$, using the Prandtl mixing length $l_P$ and the second moment of buoyancy $\overline{(g\theta_b)^2}$, was mentioned in (Burk 1981).

When $z/h \ll 1$, equations (44) take the form of (42), i.e., the surface asymptotics of the relationships of the NTLS are identical to the free convective limits of the MOST structural coefficients.

For comparison with known experimental data and the standard approximation (42), the approximations of the new theory of local similarity (44) with coefficients $v_{uu}^o = 2.2$ and $v_{\theta\theta}^o = 1.6$ are also presented in Fig. 7. According to it, the approximations of the new theory of local similarity (44) correspond to the measurement data throughout the convective layer, $0 < z/h < 1$. Whereas the standard approximation of the surface layer (42) corresponds to the measurement data only in the layer $0 \le z/h < 0.6$.

It should be emphasized that the known theories of local similarity with base parameters $gs_\theta$ and $z$ well approximate the dissipation of kinetic energy $\varepsilon_b$ and the square of buoyancy $\varepsilon_{g\theta}$ in the region $0 \le z/h < 0.55$. Therefore, approximations of the structural parameters $C_{uu}$ and $C_{\theta\theta}$ performed within the framework of classical theories of local similarity will be valid only in the lower half of the CBL.

### 7.3. Third-order structural parameters

Let $r$ be the distance along the axis $x$, and $\Delta u(x,r) = u'(x) - u'(x+r)$ be the difference in horizontal velocity at two points on axis $x$ at the same moment in time $t$. Assuming homogeneous turbulence, consider the velocity structure function of the $n$-th order:

$$D_u^{(n)}(r) = \overline{[\Delta u(r)]^n} \tag{45}$$

Consider the inertial range of the spectrum $\eta \ll r \ll \Lambda_{mw}$. Here, $\eta = \nu^{3/4} \varepsilon_b^{-1/4}$ is the Kolmogorov microscale, approximately 1 cm, and $\Lambda_{mw}$ is the integral scale of turbulence, approximately 1000 m. To describe isotropic turbulence in the inertial range of the spectrum, we accept Kolmogorov's hypotheses and use similarity theory. Then,

$$D_u^{(n)}(r) = \overline{[\Delta u(r)]^n} = C_u^{(n)} (\varepsilon_b r)^{n/3} = (C_u^{(n)} \varepsilon_b^{n/3}) r^{n/3} \tag{46}$$

where $C_u^{(n)}$ are universal constants, and $C_u^{(n)} \varepsilon_b^{n/3}$ are dimensional structural parameters of velocity of the $n$-th order, independent of $r$.

Let us put $n = 2, 3, 4$ and introduce new notations $D_{uu}(r) = D_u^{(2)}(r)$, $D_{uuu}(r) = D_u^{(3)}(r)$, $D_{uuuu}(r) = D_u^{(4)}(r)$ и $C_{uu}(r) = C_u^{(2)}(r)$, $C_{uuu}(r) = C_u^{(3)}(r)$, $C_{uuuu}(r) = C_u^{(4)}(r)$. Then, following (Atta and Chen 1970), we get:

$$\begin{cases} D_{uu}(r) = C_{uu} r^{2/3}, & C_{uu} = 4\alpha \varepsilon_b^{2/3} \\ D_{uuu}(r) = C_{uuu} r, & C_{uuu} = C_u^{(3)} \varepsilon_b \\ D_{uuuu}(r) = C_{uuuu} r^{4/3}, & C_{uuuu} = C_u^{(4)} \varepsilon_b^{4/3} \end{cases} \tag{47}$$

Assuming $\Delta g\theta_b(r) = g\theta_b(x) - g\theta_b(x+r)$ be the buoyancy difference at two points on axis $x$ at the same moment in time $t$, analogously to (45), we define the third-order structural functions, including buoyancy, $D_{\theta\theta\theta}(r)$, $D_{\theta\theta u}(r)$ and $D_{\theta u u}(r)$ such that:

$$\begin{cases} D_{\theta\theta\theta}(r) = \overline{[\Delta g\theta_b(r)]^3}, \\ D_{\theta\theta u}(r) = \overline{[\Delta g\theta_b(r)]^2 \Delta u(r)}, \\ D_{\theta u u}(r) = \overline{\Delta g\theta_b(r)[\Delta u(r)]^2} \end{cases} \tag{48}$$

At the same time, according to (Park and Atta 1980), we have:

$$D_{\theta\theta\theta}(r) = \overline{[\Delta g\theta_b(r)]^3} = 0, \qquad D_{\theta u u}(r) = \overline{\Delta g\theta_b(r)[\Delta u(r)]^2} = 0 \tag{49}$$

The relationship between third-order structural functions and second-order structural functions was first established in the studies: (Kolmogorov 1941; Yaglom 1948) see also: (Kolmogorov 1962; Monin and Yaglom 1963, 1975; Panchev and Leith 1972; Yaglom 1981).

For the structural functions of horizontal velocity and buoyancy in the region $r < \Lambda_{mw}$, Kolmogorov and Yaglom equations are:

$$6\nu \frac{d}{dr} D_{uu}(r) - D_{uuu}(r) = \frac{4}{5} \varepsilon_b r, \quad 2\chi \frac{d}{dr} D_{\theta\theta}(r) - D_{\theta\theta u}(r) = \frac{4}{3} \varepsilon_{g\theta} r \qquad (50)$$

Here, $\nu$ and $\chi$ are the coefficients of kinematic viscosity and thermal conductivity of the medium.

In the inertial range of the spectrum $\eta \ll r \ll \Lambda_{mw}$, the terms (50), including the viscosity and thermal conductivity coefficients, are very small, so:

$$-D_{uuu}(r) = \frac{4}{5} \varepsilon_b r, \quad -D_{\theta\theta u}(r) = \frac{4}{3} \varepsilon_{g\theta} r \qquad (51)$$

For the third-order structural parameters $C_{\theta\theta u}$ analogously to (46) and considering (47), we obtain:

$$C_{uuu} = -\frac{4}{5} \varepsilon_b, \quad C_{\theta\theta u} = -\frac{4}{3} \varepsilon_{g\theta} \qquad (52)$$

Relations (52) express the third-order structural parameters $C_{uuu}$ and $C_{\theta\theta u}$ through the dissipation of kinetic energy $\varepsilon_b$ and the square of buoyancy $\varepsilon_{g\theta}$. According to (52), the vertical profiles of $C_{uuu}$ and $C_{\theta\theta u}$ are geometrically similar to the profiles of $\varepsilon_b$ and $\varepsilon_{g\theta}$, see Fig. 5 and Fig. 6.

Note that the dissipation of kinetic energy and the square of buoyancy are defined by relations (25) and (29). Therefore, substitution of (25) and (29) into (52) leads to equations:

$$\begin{cases} \dfrac{C_{uuu}}{g\theta_D w_D} = -\dfrac{4}{5} \lambda^o_{\varepsilon b} \left[1 - 0.8\left(\dfrac{z}{h}\right)\right], & \lambda^o_{\varepsilon b} = 1 \\ \dfrac{C_{\theta\theta u} h}{(g\theta_D)^2 w_D} = -\dfrac{4}{3} \lambda^o_{\varepsilon g\theta} \left(\dfrac{z}{h}\right)^{-4/3} \left[1 - 0.8\left(\dfrac{z}{h}\right)\right]^{-1}, & \lambda^o_{\varepsilon g\theta} = 1/2 \end{cases} \qquad (53)$$

Equations (53) represent approximations of third-order structural parameters within the framework of local similarity theory.

## 8. Local similarity and interpretation of the Kolmogorov-Obukhov spectral theory

Let us demonstrate that the proposed theory of local similarity is consistent with the Kolmogorov-Obukhov spectral theory and the law $-5/3$. To do this, let us introduce the spectral constants of local similarity, $\alpha$ and $\beta$, assuming $4\alpha = v_{uu}^o (\lambda_{\varepsilon b}^o)^{-2/3}$, and transform equations (44) taking into account relations (25) and (29). Then:

$$\begin{cases} C_{uu} = 4\alpha \varepsilon_b^{2/3}, & C_{\theta\theta} = 4\beta \varepsilon_b^{-1/3} \varepsilon_{g\theta}, \\ 4\alpha = v_{uu}^o (\lambda_{\varepsilon b}^o)^{-2/3}, & 4\beta = v_{\theta\theta}^o (\lambda_{\varepsilon b}^o)^{1/3} (\lambda_{\varepsilon g\theta}^o)^{-1} \end{cases} \quad (54)$$

When $v_{uu}^o = 2.2$, $v_{\theta\theta}^o = 1.6$ and $\lambda_{\varepsilon b}^o = 1$, $\lambda_{\varepsilon g\theta}^o = 1/2$, the relations of the second line of (54) gives $\alpha = 0.55$ and $\beta = 0.8$.

Note that the relations in the first line of (54) can be obtained independently when determining structural parameters based on the Kolmogorov-Obukhov spectral theory and the law $-5/3$. The formulation of the spectral law $-5/3$ includes coefficients $\alpha_1$ and $\beta_1$, known as the Kolmogorov and Obukhov-Corrsin constants, respectively.

Experimental justification of the law $-5/3$ and the calculation of Kolmogorov and Obukhov-Corrsin constants were carried out, for example, in the Kansas-1968 experiment over land and presented in the papers: (Wyngaard and Coté 1971; Kaimal et al. 1972) as well as in the BOMEX-1969 and JASIN-1978 experiments over the sea and presented in the works:(Pond et al. 1971; Large and Pond 1982). According to (Wyngaard and Coté 1971; Kaimal et al. 1972), the measured coefficients are $\alpha_1 = 0.52$, $\beta_1 = 0.79$ and $\alpha_1 = 0.50$, $\beta_1 = 0.82$ respectively. While according to (Pond et al. 1971; Large and Pond 1982), the measured coefficients are $\alpha_1 = 0.55$ and $\beta_1 = 0.80$.

The correspondence of the spectral constants of local similarity, $\alpha$ and $\beta$, to the Kolmogorov and Obukhov-Corrsin constants, $\alpha_1$ and $\beta_1$, measured in completely different, independent experiments, indicates the consistency of the proposed theory of local similarity with the law $-5/3$ obtained in the Kolmogorov-Obukhov spectral theory.

## 9. Conclusion

In this article, a variant of the local similarity theory is presented, using the second moment of vertical velocity and the "spectral" Prandtl mixing length as basic parameters. This approach allows expressing the turbulent exchange coefficient, kinetic energy and buoyancy square dissipations, mixed moments of the buoyancy and vertical velocity, as well as structural parameters only through two independent basic parameters of local similarity. Comparison with experimental data convincingly confirms the correctness of the proposed approximations. Thus, the theory of local similarity significantly expands the scope of applications of the classical semi-empirical Prandtl theory and substantially complements our understanding of turbulent convection.

The proposed theory of local similarity is compared with other theories adopted for describing the atmospheric CBL. Classical similarity theory allows approximating turbulent moments with unspecified functions of dimensionless height $z/h$. However, local similarity theory is a more efficient method of research since it allows determining the analytical form of approximations of turbulent moments with an accuracy up to uncertain constants, the values of which are known from measurements in the surface layer. This position is convincingly demonstrated, for example, in the approximation of the second-order temperature structure function. Other examples of comparing the results of classical and local similarity theory are given in (Vulfson and Nikolaev 2024a).

It is established that within the framework of NTLS, approximations of the turbulent exchange coefficient and kinetic energy dissipation are fully consistent with the semi-empirical turbulence theories by Richardson (1926) and Hanna (1968) for the atmospheric CBL. The developed approximations of the structural parameters correspond to Kolmogorov-Obukhov spectral theory. Moreover, it is shown that the surface asymptotics of the local similarity theory corresponds to the free convective limits of MOST, both in the case of free convection and forced convection with weak wind. Therefore, NTLS is fully consistent with the well-known theories of turbulence under convective conditions.

It should be emphasized that the new variant of local similarity theory is more efficient for applications. Thus, known variants of local similarity theory are valid in the lower half of the CBL at $0 < z/h < 0.5$. While in the new variant, the correspondence with observational data for average turbulent parameters is observed in the layer $0 < z/h < 0.7$. For some turbulent parameters, such as kinetic energy dissipation and buoyancy square, as well as structural coefficients, the correspondence with observational data is observed throughout the convective layer, i.e., $0 < z/h < 1$.

**Acknowledgments.** is study was carried out by Vulfson A.N. within the framework of the Governmental Order to Water Problems Institute, Russian Academy of Sciences, the research project No. FMWZ-2022-0001

**Data availability statement.** Data sets generated during the current study are available from the corresponding author on reasonable request.

# References

Abdella, K., and N. Mcfarlane, 1997: A new second-order turbulence closure scheme for the planetary boundary layer. *Journal of the Atmospheric Sciences*, **54**, 1850–1867, https://doi.org/10.1175/1520-0469(1997)054<1850:ANSOTC>2.0.CO;2.

Ansmann, A., J. Fruntke, and R. Engelmann, 2010: Updraft and downdraft characterization with Doppler lidar: cloud-free versus cumuli-topped mixed layer. *Atmos. Chem. Phys.*, 14.

Atta, C. W. V., and W. Y. Chen, 1970: Structure functions of turbulence in the atmospheric boundary layer over the ocean. *J. Fluid Mech.*, **44**, 145–159, https://doi.org/10.1017/S002211207000174X.

Barenblatt, G. I., 1996: *Scaling, Self-similarity, and Intermediate Asymptotics*. Cambridge University Press,.

Batchelor, G. K., 1950: The application of the similarity theory of turbulence to atmospheric diffusion. *Quart J Royal Meteoro Soc*, **76**, 133–146, https://doi.org/10.1002/qj.49707632804.

——, 1953: The conditions for dynamical similarity of motions of a frictionless perfect-gas atmosphere. *Quart J Royal Meteoro Soc*, **79**, 224–235, https://doi.org/10.1002/qj.49707934004.

Berg, L. K., R. K. Newsom, and D. D. Turner, 2017: Year-Long Vertical Velocity Statistics Derived from Doppler Lidar Data for the Continental Convective Boundary Layer. *Journal of Applied Meteorology and Climatology*, **56**, 2441–2454, https://doi.org/10.1175/JAMC-D-16-0359.1.

Bernard-Trottolo, S., B. Campistron, A. Druilhet, F. Lohou, and F. Saïd, 2004: TRAC98: Detection of Coherent Structures in a Convective Boundary Layer using Airborne Measurements. *Boundary-Layer Meteorology*, **111**, 181–224, https://doi.org/10.1023/B:BOUN.0000016465.50697.63.

Beyrich, F., and Coauthors, 2012: Towards a Validation of Scintillometer Measurements: The LITFASS-2009 Experiment. *Boundary-Layer Meteorol*, **144**, 83–112, https://doi.org/10.1007/s10546-012-9715-8.

Boers, R., 1989: A Parameterization of the Depth of the Entrainment Zone. *J. Appl. Meteor.*, **28**, 107–111, https://doi.org/10.1175/1520-0450(1989)028<0107:APOTDO>2.0.CO;2.

Boussinesq, J., 1870: Essai théorique sur les lois trouvées expérimentalement par M. Bazin pour l'écoulement uniforme de l'eau dans les canaux découverts. *CR Acad. Sci. Paris*, **71**, 389–393.

Buckingham, E., 1914: On physically similar systems; Illustrations of the use of dimensional equations. *Physical Review*, **4**, 345–376, https://doi.org/10.1103/PhysRev.4.345.

Burchard, H., and O. Petersen, 1999: Models of turbulence in the marine environment —a comparative study of two-equation turbulence models. *Journal of Marine Systems*, **21**, 29–53, https://doi.org/10.1016/S0924-7963(99)00004-4.

Burk, S. D., 1981: Comparison of Structure Parameter Scaling Expressions with Turbulence Closure Model Predictions. *Journal of the Atmospheric Sciences*, **38**, 751–761, https://doi.org/10.1175/1520-0469(1981)038<0751:COSPSE>2.0.CO;2.

Calder, K. L., 1949: The criterion of turbulence in a fluid of variable density, with particular reference to conditions in the atmosphere. *Quart J Royal Meteoro Soc*, **75**, 71–88, https://doi.org/10.1002/qj.49707532311.

Caughey, S. J., and S. G. Palmer, 1979: Some aspects of turbulence structure through the depth of the convective boundary layer. *Quarterly Journal of the Royal Meteorological Society*, **105**, 811–827, https://doi.org/10.1002/qj.49710544606.

Deardorff, J. W., 1970: Convective Velocity and Temperature Scales for the Unstable Planetary Boundary Layer and for Rayleigh Convection. *Journal of the Atmospheric Sciences*, **27**, 1211–1213, https://doi.org/10.1175/1520-0469(1970)027<1211:cvatsf>2.0.co;2.

Degrazia, G. A., D. M. Moreira, and M. T. Vilhena, 2001: Derivation of an Eddy Diffusivity Depending on Source Distance for Vertically Inhomogeneous Turbulence in a Convective Boundary Layer. *Journal of Applied Meteorology*, **40**, 1233–1240, https://doi.org/10.1175/1520-0450(2001)040<1233:DOAEDD>2.0.CO;2.

Degrazia, G. A., and Coauthors, 2015: Eddy diffusivities for the convective boundary layer derived from LES spectral data. *Atmospheric Pollution Research*, **6**, 605–611, https://doi.org/10.5094/APR.2015.068.

Dosio, A., J. V. Guerau de Arellano, A. A. M. Holtslag, and P. J. H. Builtjes, 2005: Relating Eulerian and Lagrangian Statistics for the Turbulent Dispersion in the Atmospheric Convective Boundary Layer. *Journal of the Atmospheric Sciences*, **62**, 1175–1191, https://doi.org/10.1175/JAS3393.1.

Druilhet, A., J. P. Frangi, D. Guedalia, and J. Fontan, 1983: Experimental Studies of the Turbulence Structure Parameters of the Convective Boundary Layer. *Journal of Climate and Applied Meteorology*, **22**, 594–608, https://doi.org/10.1175/1520-0450(1983)022<0594:ESOTTS>2.0.CO;2.

Fedorovich, E., J. A. Gibbs, and A. Shapiro, 2017: Numerical Study of Nocturnal Low-Level Jets over Gently Sloping Terrain. *Journal of the Atmospheric Sciences*, **74**, 2813–2834, https://doi.org/10.1175/JAS-D-17-0013.1.

Fedorovich, E. E., and D. V. Mironov, 1995: A model for a shear-free convective boundary layer with parameterized capping inversion structure. *Journal of the Atmospheric Sciences*, **52**, 83–95, https://doi.org/10.1175/1520-0469(1995)052<0083:amfasf>2.0.co;2.

Ferrero, E., N. M. Colonna, and U. Rizza, 2009: Non-local simulation of the stable boundary layer with a third order moments closure model. *Journal of Marine Systems*, **77**, 495–501, https://doi.org/10.1016/j.jmarsys.2008.11.013.

Fischer, H. B., 1979: *Mixing in inland and coastal waters*. Academic press,.

Fodor, K., and J. P. Mellado, 2020: New Insights into Wind Shear Effects on Entrainment in Convective Boundary Layers Using Conditional Analysis. *Journal of the Atmospheric Sciences*, **77**, 3227–3248, https://doi.org/10.1175/JAS-D-19-0345.1.

Frisch, A. S., and G. R. Ochs, 1975: A Note on the Behavior of the Temperature Structure Parameter in a Convective Layer Capped by a Marine Inversion. *J. Appl. Meteor.*, **14**, 415–419, https://doi.org/10.1175/1520-0450(1975)014<0415:ANOTBO>2.0.CO;2.

Frisch, U., 1995: *Turbulence: The Legacy of A. N. Kolmogorov*. Cambridge University Press,.

Garcia, J. R., 2014: Analysis of the surface layer and the entrainment zone of a convective boundary layer using direct numerical simulation. *Reports on Earth System Science*, **159**.

Gentine, P., A. K. Betts, B. R. Lintner, K. L. Findell, C. C. van Heerwaarden, A. Tzella, and F. D'Andrea, 2013: A Probabilistic Bulk Model of Coupled Mixed Layer and Convection. Part I: Clear-Sky Case. *Journal of the Atmospheric Sciences*, **70**, 1543–1556, https://doi.org/10.1175/JAS-D-12-0145.1.

George, W. K., 2013: Lectures in Turbulence for the 21st Century. *Chalmers University of Technology*, **550**.

Gibbs, J. A., and E. Fedorovich, 2014: Comparison of Convective Boundary Layer Velocity Spectra Retrieved from Large- Eddy-Simulation and Weather Research and Forecasting Model Data. *Journal of Applied Meteorology and Climatology*, **53**, 377–394, https://doi.org/10.1175/JAMC-D-13-033.1.

——, and ——, 2020: On the Evaluation of the Proportionality Coefficient between the Turbulence Temperature Spectrum and Structure Parameter. *Journal of the Atmospheric Sciences*, **77**, 2761–2763, https://doi.org/10.1175/JAS-D-19-0344.1.

——, ——, B. Maronga, C. Wainwright, and M. Dröse, 2016: Comparison of Direct and Spectral Methods for Evaluation of the Temperature Structure Parameter in Numerically Simulated Convective Boundary Layer Flows. *Monthly Weather Review*, **144**, 2205–2214, https://doi.org/10.1175/MWR-D-15-0390.1.

Goulart, A. G., D. M. Moreira, J. C. Carvalho, and T. Tirabassi, 2004: Derivation of eddy diffusivities from an unsteady turbulence spectrum. *Atmospheric Environment*, **38**, 6121–6124, https://doi.org/10.1016/j.atmosenv.2004.08.010.

Greenhut, G. K., and G. Mastrantonio, 1989: Turbulence Kinetic Energy Budget Profiles Derived from Doppler Sodar Measurements. *Journal of Applied Meteorology*, **28**, 99–106, https://doi.org/10.1175/1520-0450(1989)028<0099:TKEBPD>2.0.CO;2.

Gryanik, V. M., and J. Hartmann, 2002: A Turbulence Closure for the Convective Boundary Layer Based on a Two-Scale Mass-Flux Approach. *Journal of the Atmospheric Sciences*, **59**, 2729–2744, https://doi.org/10.1175/1520-0469(2002)059<2729:ATCFTC>2.0.CO;2.

Guillemet, B., H. Isaka, and P. Mascart, 1983: Molecular dissipation of turbulent fluctuations in the convective mixed layer part I: Height variations of dissipation rates. *Boundary-Layer Meteorology*, **27**, 141–162, https://doi.org/10.1007/BF00239611.

Hanna, S. R., 1968: A Method of Estimating Vertical Eddy Transport in the Planetary Boundary Layer Using Characteristics of the Vertical Velocity Spectrum. *Journal of the Atmospheric Sciences*, **25**, 1026–1033, https://doi.org/10.1175/1520-0469(1968)025<1026:AMOEVE>2.0.CO;2.

——, 1984: Applications in Air Pollution Modeling. *Atmospheric Turbulence and Air Pollution Modelling*, F.T.M. Nieuwstadt and H. Dop, Eds., Springer Netherlands, 275–310.

Hildebrand, P. H., and B. Ackerman, 1984: Urban Effects on the Convective Boundary Layer. *Journal of the Atmospheric Sciences*, **41**, 76–91, https://doi.org/10.1175/1520-0469(1984)041<0076:UEOTCB>2.0.CO;2.

Hinze, J. O., 1975: *Turbulence*. McGraw-Hill,.

Holt, T., and S. Raman, 1988: A review and comparative evaluation of multilevel boundary layer parameterizations for first-order and turbulent kinetic energy closure schemes. *Reviews of Geophysics*, **26**, 761–780, https://doi.org/10.1029/RG026i004p00761.

Holtslag, A. A. M., and C.-H. Moeng, 1991: Eddy Diffusivity and Countergradient Transport in the Convective Atmospheric Boundary Layer. *Journal of the Atmospheric Sciences*, **48**, 1690–1698, https://doi.org/10.1175/1520-0469(1991)048<1690:EDACTI>2.0.CO;2.

Isihara, A., 1971: *Statistical physics*. Academic Press, 454 pp.

Kader, B. A., and A. M. Yaglom, 1990: Mean fields and fluctuation moments in unstably stratified turbulent boundary layers. *Journal of Fluid Mechanics*, **212**, 637–662, https://doi.org/10.1017/S0022112090002129.

Kaimal, J. C., 1973: Turbulenece spectra, length scales and structure parameters in the stable surface layer. *Boundary-Layer Meteorology*, **4**, 289–309, https://doi.org/10.1007/BF02265239.

——, and D. A. Haugen, 1967: Characteristics of vertical velocity fluctuations observed on a 430-m tower. *Q.J Royal Met. Soc.*, **93**, 305–317, https://doi.org/10.1002/qj.49709339703.

Kaimal, J. C., and J. J. Finnigan, 1994: *Atmospheric boundary layer flows: their structure and measurement*. Oxford university press,.

Kaimal, J. C., J. C. Wyngaard, Y. Izumi, and O. R. Coté, 1972: Spectral characteristics of surface-layer turbulence. *Quarterly Journal of the Royal Meteorological Society*, **98**, 563–589, https://doi.org/10.1002/qj.49709841707.

——, ——, D. A. Haugen, O. R. Cote, Y. Izumi, S. J. Caughey, and C. J. Readings, 1976: Turbulence Structure in the Convective Boundary Layer. *Journal of the Atmospheric Sciences*, **33**, 2152–2169, https://doi.org/10.1175/1520-0469(1976)033<2152:TSITCB>2.0.CO;2.

Kang, S. L., K. J. Davis, and M. LeMone, 2007: Observations of the ABL structures over a heterogeneous land surface during IHOP_2002. *Journal of Hydrometeorology*, **8**, 221–244, https://doi.org/10.1175/JHM567.1.

von Kármán, T., 1930: Mechanische Ahnlichkeit und Turbulenz Nach. *Ges. Wiss. Gottingen, Math.-Phys*, 58–76.

Kolmogorov, A. N., 1941: Dissipation of energy in locally isotropic turbulence. *CR Dokl Acad Sci USSR*, Vol. 32 of, 16–18.

——, 1942: Equations of turbulent motion of an incompressible fluid. *Izvestiya AN SSSR. Ser. fiz.*, **6**, 56–58.

Kolmogorov, A. N., 1962: A refinement of previous hypotheses concerning the local structure of turbulence in a viscous incompressible fluid at high Reynolds number. *J. Fluid Mech.*, **13**, 82–85, https://doi.org/10.1017/S0022112062000518.

Kolmogorov, A. N., 1968: Local structure of turbulence in an incompressible viscous fluid at vary hight Reinolds numbers. *Sov. Phys. Usp.*, **10**, 734–746, https://doi.org/10.1070/PU1968v010n06ABEH003710.

Kramm, G., D. J. Amaya, T. Foken, and N. Mölders, 2013: Hans A. Panofsky's Integral Similarity Function—At Fifty. *Atmospheric and Climate Sciences*, **03**, 581–594, https://doi.org/10.4236/acs.2013.34061.

Large, W. G., and S. Pond, 1982: Sensible and Latent Heat Flux Measurements over the Ocean. *J. Phys. Oceanogr.*, **12**, 464–482, https://doi.org/10.1175/1520-0485(1982)012<0464:SALHFM>2.0.CO;2.

LeMone, M. A., M. Tewari, F. Chen, and J. Dudhia, 2013: Objectively Determined Fair-Weather CBL Depths in the ARW-WRF Model and Their Comparison to CASES-97 Observations. *Monthly Weather Review*, **141**, 30–54, https://doi.org/10.1175/MWR-D-12-00106.1.

Lenschow, D. H., and P. L. Stephens, 1980: The role of thermals in the convective boundary layer. *Boundary-Layer Meteorology*, **19**, 509–532, https://doi.org/10.1007/BF00122351.

Lenschow, D. H., and B. B. Stankov, 1986: Length Scales in the Convective Boundary Layer. *J. Atmos. Sci.*, **43**, 1198–1209, https://doi.org/10.1175/1520-0469(1986)043<1198:LSITCB>2.0.CO;2.

Lenschow, D. H., J. C. Wyngaard, and W. T. Pennell, 1980: Mean-field and second-moment budgets in a baroclinic, convective boundary layer. *Journal of the Atmospheric Sciences*, **37**, 1313–1326, https://doi.org/10.1175/1520-0469(1980)037<1313:MFASMB>2.0.CO;2.

Lenschow, D. H., M. Lothon, S. D. Mayor, P. P. Sullivan, and G. Canut, 2012: A Comparison of Higher-Order Vertical Velocity Moments in the Convective Boundary Layer from Lidar with In Situ Measurements and Large-Eddy Simulation. *Boundary-Layer Meteorology*, **143**, 107–123, https://doi.org/10.1007/s10546-011-9615-3.

Lothon, M., D. H. Lenschow, and S. D. Mayor, 2009: Doppler Lidar Measurements of Vertical Velocity Spectra in the Convective Planetary Boundary Layer. *Boundary-Layer Meteorology*, **132**, 205–226, https://doi.org/10.1007/s10546-009-9398-y.

Luce, H., L. Kantha, H. Hashiguchi, A. Doddi, D. Lawrence, and M. Yabuki, 2020: On the Relationship between the TKE Dissipation Rate and the Temperature Structure Function Parameter in the Convective Boundary Layer. *Journal of the Atmospheric Sciences*, **77**, 2311–2326, https://doi.org/10.1175/JAS-D-19-0274.1.

Mahrt, L., 1986: On the Shallow Motion Approximations. *Journal of Atmospheric Sciences*, **43**, 1036–1044, https://doi.org/10.1175/1520-0469(1986)043<1036:OTSMA>2.0.CO;2.

McColl, K. A., C. C. van Heerwaarden, G. G. Katul, P. Gentine, and D. Entekhabi, 2017: Role of large eddies in the breakdown of the Reynolds analogy in an idealized mildly unstable atmospheric surface layer. *Quarterly Journal of the Royal Meteorological Society*, **143**, 2182–2197, https://doi.org/10.1002/qj.3077.

Mellor, G. L., 1973: Analytic Prediction of the Properties of Stratified Planetary Surface Layers. *J. Atmos. Sci.*, **30**, 1061–1069, https://doi.org/10.1175/1520-0469(1973)030<1061:APOTPO>2.0.CO;2.

Menter, F. R., and Y. Egorov, 2010: The Scale-Adaptive Simulation Method for Unsteady Turbulent Flow Predictions. Part 1: Theory and Model Description. *Flow Turbulence Combust*, **85**, 113–138, https://doi.org/10.1007/s10494-010-9264-5.

Mihaljan, J. M., 1962: A Rigorous Exposition of the Boussinesq Approximations Applicable to a Thin Layer of Fluid. *The Astrophysical Journal*, **136**, 1126, https://doi.org/10.1086/147463.

Monin, A. S., 1958: The Structure of Atmospheric Turbulence. *Theory Probab. Appl.*, **3**, 266–296, https://doi.org/10.1137/1103023.

Monin, A. S., and A. M. Obukhov, 1953: Dimensionless characteristics of turbulence in the atmospheric surface layer. *Doklady Akademii Nauk SSSR*, **93**, 253–226.

——, and ——, 1954: Basic laws of turbulent mixing in the surface layer of the atmosphere. *Contrib. Geophys. Inst. Acad. Sci. USSR*, **24**, 163–187.

——, and A. M. Yaglom, 1963: ON THE LAWS OF SMALL-SCALE TURBULENT FLOW OF LIQUIDS AND GASES. *Russ. Math. Surv.*, **18**, 89–109, https://doi.org/10.1070/RM1963v018n05ABEH004133.

——, and ——, 1975: Mechanics of turbulence. *Statistical Fluid Mechanics*, **2**.

Noh, Y., W. G. Cheon, S. Y. Hong, and S. Raasch, 2003: Improvement of the K-profile Model for the Planetary Boundary Layer based on Large Eddy Simulation Data. *Boundary-Layer Meteorology*, **107**, 401–427, https://doi.org/10.1023/A:1022146015946.

Obukhov, A. M., 1946: Turbulence in a temperature-inhomogeneous atmosphere. *Tr. Institute of theory. geophysicist of the Academy of Sciences of the USSR*, **1**, 95–115.

——, 1949: The structure of the temperature field in a turbulent flow. *Изв. АН СССР. Сер. геогр. и геофиз*, **13**, 58–69.

——, 1959: On the influence of Archimedean forces on the structure of the temperature field in a turbulent flow. *Doklady Akademii Nauk SSSR*, **125**, 1246.

Ogura, Y., 1952: The Theory of Turbulent Diffusion in the Atmosphere (I). *Journal of the Meteorological Society of Japan*, **30**, 23–28, https://doi.org/10.2151/jmsj1923.30.1_23.

Ogura, Y., and N. A. Phillips, 1962: Scale Analysis of Deep and Shallow Convection in the Atmosphere. *Journal of the Atmospheric Sciences*, **19**, 173–179, https://doi.org/10.1175/1520-0469(1962)019<0173:saodas>2.0.co;2.

Olver, P. J., 1993: *Applications of Lie Groups to Differential Equations*. Springer US,.

Onsager, L., 1949: Statistical hydrodynamics. *Nuovo Cim*, **6**, 279–287, https://doi.org/10.1007/BF02780991.

Panchev, S., and C. E. Leith, 1972: Random Functions and Turbulence. *American Journal of Physics*, **40**, 927–927, https://doi.org/10.1119/1.1986709.

Panofsky, H. A., 1978: Matching in the Convective Planetary Boundary Layer. *Journal of the Atmospheric Sciences*, **35**, 272–276, https://doi.org/10.1175/1520-0469(1978)035<0272:MITCPB>2.0.CO;2.

Park, J. T., and C. W. Van Atta, 1980: Hot- and cold-wire sensitivity corrections for moments of the fine scale turbulence in heated flows. *Phys. Fluids*, **23**, 701, https://doi.org/10.1063/1.863056.

Peña, A., S.-E. E. Gryning, J. Mann, and C. B. Hasager, 2010: Length scales of the neutral wind profile over homogeneous terrain. *Journal of Applied Meteorology and Climatology*, **49**, 792–806, https://doi.org/10.1175/2009JAMC2148.1.

Pond, S., G. T. Phelps, J. E. Paquin, G. McBean, and R. W. Stewart, 1971: Measurements of the Turbulent Fluxes of Momentum, Moisture and Sensible Heat over the Ocean. *J. Atmos. Sci.*, **28**, 901–917, https://doi.org/10.1175/1520-0469(1971)028<0901:MOTTFO>2.0.CO;2.

Prandtl, L., 1925: 7. Bericht über Untersuchungen zur ausgebildeten Turbulenz. *ZAMM - Journal of Applied Mathematics and Mechanics / Zeitschrift für Angewandte Mathematik und Mechanik*, **5**, 136–139, https://doi.org/10.1002/zamm.19250050212.

——, 1932: Meteorogische anwendung der stromungslehre. *Beitr. Phys. fr. Atmoshare.*, **19**, 188–202.

——, 1945: *Uber ein neues Formelsystem fur die ausgebildete Turbulenz*, Nachrichten der Akademie der Wissenschaften zu Gottingen. Mathematish- Physikalische. Klasse., 6–19.

Richardson, L., 1926: Atmospheric diffusion shown on a distance-neighbour graph. *Proc. R. Soc. Lond. A*, **110**, 709–737, https://doi.org/10.1098/rspa.1926.0043.

Roberts, P., and D. Webster, 2002: Turbulent diffusion in environmental fluid mechanics theories and application. *ASCE, Reston, USA*, 1–42.

Rytov, S. M., I. A. Kravt͡sov, and V. I. Tatarskiĭ, 1989: *Principles of statistical radiophysics: Wave propagation through random media*. Springer,.

Shao, Y., J. M. Hacker, and P. Schwerdtfeger, 1991: The structure of turbulence in a coastal atmospheric boundary layer. *Quarterly Journal of the Royal Meteorological Society*, **117**, 1299–1324, https://doi.org/10.1002/qj.49711750209.

Sommeria, G., and M. A. LeMone, 1978: Direct Testing of a Three-Dimensional Model of the Planetary Boundary Layer Against Experimental Data. *Journal of the Atmospheric Sciences*, **35**, 25–39, https://doi.org/10.1175/1520-0469(1978)035<0025:DTOATD>2.0.CO;2.

Sorbjan, Z., 1986: On similarity in the atmospheric boundary layer. *Boundary-Layer Meteorology*, **34**, 377–397, https://doi.org/10.1007/BF00120989.

——, 1987: An examination of local similarity theory in the stably stratified boundary layer. *Boundary-Layer Meteorology*, **38**, 63–71, https://doi.org/10.1007/BF00121555.

——, 1988: Local similarity in the convective boundary layer (CBL). *Boundary-Layer Meteorology*, **45**, 237–250, https://doi.org/10.1007/BF01066672.

——, 1990: Similarity Scales and Universal Profiles of Statistical Moments in the Convective Boundary Layer. *Journal of Applied Meteorology*, **29**, 762–775, https://doi.org/10.1175/1520-0450(1990)029<0762:SSAUPO>2.0.CO;2.

——, 1991: Evaluation of Local Similarity Functions in the Convective Boundary Layer. *Journal of Applied Meteorology*, **30**, 1565–1583, https://doi.org/10.1175/1520-0450(1991)030<1565:EOLSFI>2.0.CO;2.

Spiegel, E. A., and G. Veronis, 1960: On the Boussinesq Approximation for a Compressible Fluid. *The Astrophysical Journal*, **131**, 442, https://doi.org/10.1086/146849.

Sullivan, P. P., C.-H. Moeng, B. Stevens, D. H. Lenschow, and S. D. Mayor, 1998: Structure of the Entrainment Zone Capping the Convective Atmospheric Boundary Layer. *J. Atmos. Sci.*, **55**, 3042–3064, https://doi.org/10.1175/1520-0469(1998)055<3042:SOTEZC>2.0.CO;2.

Sun, W.-Y., 1986: Air pollution in a convective boundary layer. *Atmospheric Environment (1967)*, **20**, 1877–1886, https://doi.org/10.1016/0004-6981(86)90328-8.

——, 1989: Numerical study of dispersion in the convective boundary layer. *Atmospheric Environment (1967)*, **23**, 1205–1217, https://doi.org/10.1016/0004-6981(89)90147-9.

Tatarski, V. I., R. A. Silverman, and N. Chako, 1961: Wave Propagation in a Turbulent Medium. *Physics Today*, **14**, 46–51, https://doi.org/10.1063/1.3057286.

Tsvang, L. R., 1969: Microstructure of Temperature Fields in the Free Atmosphere. *Radio Science*, **4**, 1175–1177, https://doi.org/10.1029/RS004i012p01175.

Turner, J. S., 1962: The 'starting plume' in neutral surroundings. *Journal of Fluid Mechanics*, **13**, 356–368, https://doi.org/10.1017/S0022112062000762.

Vallis, G. K., 2017: *Atmospheric and oceanic fluid dynamics*. Cambridge University Press,.

Vulfson, A. N., 1981: Equations of Deep Convection in a Dry Atmosphere. *Izvestiya. Atmospheric and Oceanic Physics*, **17**, 646–649.

Vulfson, A. N., and P. V. Nikolaev, 2022: Local Similarity Theory of Convective Turbulent Layer Using "Spectral" Prandtl Mixing Length and Second Moment of Vertical Velocity. *Journal of the Atmospheric Sciences*, **79**, 101–118, https://doi.org/10.1175/JAS-D-20-0330.1.

Vulfson, A. N., and P. V. Nikolaev, 2024a: Classical and local similarity in problems of turbulent convection: Extension of Prandtl semi-empirical theory for horizontal layers of water and air mediums. *Physics of Fluids*, **36**, 026612, https://doi.org/10.1063/5.0176848.

——, and ——, 2024b: Variant of the Local Similarity Theory and Approximations of Vertical Profiles of Turbulent Moments of the Atmospheric Convective Boundary Layer. *Izv. Atmos. Ocean. Phys.*, **60**, 48–58, https://doi.org/10.1134/S0001433824700038.

——, I. A. Volodin, and O. O. Borodin, 2004: Local theory of similarity and universal profiles of turbulent characteristics of a connective boundary layer. *Meteorologiya i Gidrologiya*, **10**, 5–15.

Vulfson, N. I., 1964: *Convective Motions in a Free Atmosphere*.

Wamser, C., H. Moller, and H. A. Panofsky, 1977: On the spectral scale of wind fluctuations within and above the surface layer. *Quarterly Journal of the Royal Meteorological Society*, **103**, 721–730, https://doi.org/10.1002/qj.49710343814.

Weill, A., C. Klapisz, B. Strauss, F. Baudin, C. Jaupart, P. Van Grunderbeeck, and J. P. Goutorbe, 1980: Measuring Heat Flux and Structure Functions of Temperature Fluctuations with an Acoustic Doppler Sodar. *Journal of Applied Meteorology*, **19**, 199–205, https://doi.org/10.1175/1520-0450(1980)019<0199:MHFASF>2.0.CO;2.

Weiss, A., 2002: Determination of thermal stratification and turbulence of the atmospheric surface layer over various types of terrain by optical scientillometry.

Wood, C. R., and Coauthors, 2010: Turbulent Flow at 190 m Height Above London During 2006-2008: A Climatology and the Applicability of Similarity Theory. *Boundary-Layer Meteorology*, **137**, 77–96, https://doi.org/10.1007/s10546-010-9516-x.

Wyngaard, J. C., and O. R. Coté, 1971: The Budgets of Turbulent Kinetic Energy and Temperature Variance in the Atmospheric Surface Layer. *Journal of the Atmospheric Sciences*, **28**, 190–201, https://doi.org/10.1175/1520-0469(1971)028<0190:TBOTKE>2.0.CO;2.

——, and S. F. Clifford, 1978: Estimating Momentum, Heat and Moisture Fluxes from Structure Parameters. *J. Atmos. Sci.*, **35**, 1204–1211, https://doi.org/10.1175/1520-0469(1978)035<1204:EMHAMF>2.0.CO;2.

——, and M. A. LeMone, 1980: Behavior of the Refractive Index Structure Parameter in the Entraining Convective Boundary Layer. *Journal of the Atmospheric Sciences*, **37**, 1573–1585, https://doi.org/10.1175/1520-0469(1980)037<1573:BOTRIS>2.0.CO;2.

——, Y. Izumi, and S. A. Collins, 1971: Behavior of the Refractive-Index-Structure Parameter near the Ground*. *Journal of the Optical Society of America*, **61**, 1646, https://doi.org/10.1364/JOSA.61.001646.

——, W. T. Pennell, D. H. Lenschow, and M. A. LeMone, 1978: The Temperature-Humidity Covariance Budget in the Convective Boundary Layer. *Journal of the Atmospheric Sciences*, **35**, 47–58, https://doi.org/10.1175/1520-0469(1978)035<0047:TTHCBI>2.0.CO;2.

Yaglom, A., 1981: Laws of small scale turbulence in atmosphere and ocean (in commemoration of the 40th anniversary of the theory of locally isotropic turbulence). *Izv. Atmos. Oceanic Phys.*, **17**, 1235–1257.

Yaglom, A. M., 1948: Homogeneous and isotropic turbulence in a viscous compressible fluid. *Izv. Akad. Nauk. SSSR Ser. Geogr. i Geofiz*, **12**, 501.

Young, G. S., 1988: Turbulence structure of the convective boundary layer. Part I: variability of normalized turbulence statistics. *Journal of the Atmospheric Sciences*, **45**, 719–726, https://doi.org/10.1175/1520-0469(1988)045<0719:TSOTCB>2.0.CO;2.

Zeman, O., and J. L. Lumley, 1976: Modeling Buoyancy Driven Mixed Layers. *Journal of the Atmospheric Sciences*, **33**, 1974–1988, https://doi.org/10.1175/1520-0469(1976)033<1974:MBDML>2.0.CO;2.

Zhou, B., J. S. Simon, and F. K. Chow, 2014: The Convective Boundary Layer in the Terra Incognita. *Journal of the Atmospheric Sciences*, **71**, 2545–2563, https://doi.org/10.1175/JAS-D-13-0356.1.